\documentclass[10pt, conference, compsocconf]{IEEEtran}
\usepackage{cite}
\usepackage{amsmath,amssymb,amsfonts}
\usepackage{algorithmic}
\usepackage{textcomp}
\usepackage{xcolor}
\def\BibTeX{{\rm B\kern-.05em{\sc i\kern-.025em b}\kern-.08em
    T\kern-.1667em\lower.7ex\hbox{E}\kern-.125emX}}
\usepackage{blindtext, graphicx}
\usepackage{caption}
\usepackage{subcaption}
\usepackage{array}
\let\labelindent\relax
\usepackage{enumitem}
\usepackage{multirow}
\usepackage{footnote}
\usepackage{url}
\usepackage{todonotes}
\usepackage{makecell}
\usepackage{comment}
\newcommand\tbt[1]{\scriptsize{#1}}
\newcommand{\cmmnt}[1]{\ignorespaces}

\begin{document}

\bstctlcite{IEEEexample:BSTcontrol}

\title{Experimental Resilience Assessment of \\An Open-Source Driving Agent\vspace{-2ex}}

\author{\IEEEauthorblockN{Abu Hasnat Mohammad Rubaiyat, Yongming Qin, Homa Alemzadeh}
\IEEEauthorblockA{Department of Electrical and Computer Engineering, University of Virginia\\
\{ar3fx, yq9vc, ha4d\}@virginia.edu}\\
}

\maketitle
\begin{abstract}
Autonomous vehicles (AV) depend on the sensors like RADAR and camera for the perception of the environment, path planning, and control. With the increasing autonomy and interactions with the complex environment, there have been growing concerns regarding the safety and reliability of AVs. This paper presents a Systems-Theoretic Process Analysis (STPA) based fault injection framework to assess the resilience of an open-source driving agent, called \emph{openpilot}, under different environmental conditions and faults affecting sensor data. To increase the coverage of unsafe scenarios during testing, we use a strategic software fault-injection approach where the triggers for injecting the faults are derived from the unsafe scenarios identified during the high-level hazard analysis of the system. The experimental results show that the proposed strategic fault injection approach increases the hazard coverage compared to random fault injection and, thus, can help with more effective simulation of safety-critical faults and testing of AVs. In addition, the paper provides insights on the performance of \emph{openpilot} safety mechanisms and its ability in timely detection and recovery from faulty inputs.
%the environmental irregularities can degrade the performance of the existing autonomous control systems of openpilot, and sometimes can create hazardous scenarios that may result in fatal accidents.
\end{abstract}

% \fi
\IEEEpeerreviewmaketitle
\section{Introduction}
%- What is the problem? Why is this important?
Autonomous vehicles are one of the most complex software-intensive cyber-physical systems (CPS). In addition to the basic car mechanisms (e.g. gas/brake system, steering system), they are equipped with driving assistance mechanisms such as Adaptive Cruise Control (ACC), Lane Keeping Assist System (LKAS), and Assisted Lane Change. AVs use smart sensors (e.g. camera, RADAR, LIDAR) and machine learning (ML) algorithms for perception of the surrounding environment, path finding, and navigation. For example, LKAS uses computer vision and ML to process camera data, locate the right/left lane markers, and adjust the steering angle to keep the vehicle inside the lanes.  %system is currently one of the most visited research areas in the field of artificial intelligence. Researchers from different corners of the world are working to introduce complete autonomy to the automotive industry. 
%Driving assistance systems such as Adaptive Cruise Control (ACC), Lane Keeping Assist System (LKAS), Assisted Lane Change are already being applied to the latest commercial car models. 

With increasing deployment of AVs on the road and the goal of moving towards full autonomy in near future, reliability and safety of these systems are of high concern. There have been already several reports on incidents involving AVs, e.g., the fatal crashes of Tesla model S in 2016 \cite{TeslaIncident2016}, model X in 2018 \cite{TeslaIncident2018}, and Uber accident in 2018 \cite{UberIncident}. 
%\todo[author=HA,inline]{Reference to real incident examples with lead crash, out of lane, and sudden stop due to faults in sensor data. Deeptest has some examples., e.g., the recent Uber Incident \cite{UberIncident}--DONE}...
The safe operation of an AV depends not only on the proper functioning of the sensors, actuators, and mechanical components, but also the proper operation of the autonomous control software and its interactions with other components, human driver, and environment. The faults in the controller may become activated either by the environmental conditions or human errors, and get propagated into the system resulting in safety hazards. 

Current practice in testing and safety validation of AVs involves simulation testing using realistic scenarios in virtual environments or real-world road testing. However, given the high costs of developing testbeds and running experiments, a question that arises is what constitutes an adequate testing experiment or how much testing is good enough? The safety-critical faults causing hazards and incidents are rare and it might take forever for an AV to experience them during road test experiments. To ensure resilience against safety-critical faults, we need techniques for specifically testing the influence of such faults using fault injection. 
% The hazard we consider is different from theirs.
%no previous work studied the resilience of real control software of AVs and its ability in detection and mitigation of faults.

Most of the previous works in this area focus on analysis of adversarial attacks on ML systems and assessment of resilience of ML algorithms used for computer vision. Evtimov et al. \cite{evtimov2017robust} proposed Robust Physical Perturbations (RP2), a robust attack algorithm that generates physically realizable adversarial perturbations of the road signs by using perturbation masks in the shape of graffiti. DeepXplore \cite{Pei:Deepxplore} and DeepTest \cite{DeepTest:Tian} focused on maximizing neuron coverage to generate perturbed images to test deep learning based algorithms. 
%\todo[inline]{write about adversarial ML and how they try to solve the problem by finding the corner cases for ML}. 
Others have studied attacks to the AV's radar module, such as creating a ghost vehicle or radar jamming \cite{RadarFault:Petit}. %Many researchers are also working on the defense mechanisms against the attacks on communication network of connected vehicles \cite{alheeti2015intrusion} \cite{garip2015congestion}. 
%\todo[inline]{cite works on attacks to radar and the attacks to platoon of cars}
However, an open question is whether the propagation of such faults into the control software could actually lead to unsafe scenarios and incidents.

This paper studies the resilience of an open-source driving agent, called \textit{openpilot} \cite{openpilot:Git}, which has been used in some car models on real roads, against faults and environmental conditions affecting the sensor data, including RADAR, camera, and car sensors for steering angle and speed. We specifically focus on these faults as they directly affect the performance of ML and perception systems, which are reported to be a major cause ($\sim$44\%) of disengagement incidents in AVs \cite{subho2018hands}. We have created a simulation environment consisting of \textit{openpilot} control software integrated with a computer vision module for real-time processing of the previously collected video data captured from the roads and a fault injection framework that mimics the effect of sensor faults and environmental conditions. We study the propagation of faults into the system and assess the ability of the control software in masking the faults, timely detection of errors and raising alerts, and mitigating safety hazards. %Although previous work shows preliminary results of fault injection into AVs \cite{jha2018avfi}, we consider a more realistic AV control software that is already being deployed in real cars used on the road. Compared to \cite{jha2018avfi}, we consider several environmental conditions and faults affecting the sensor data and different types of hazards to assess the system's resilience. STPA analysis is used to find the locations of injecting faults or affecting factors that matter a lot.\todo{Revise}

To reduce the test space and increase the probability of generation of unsafe scenarios, we present a \textit{strategic} software fault injection approach where the triggers for injecting the faults are driven by the high-level hazard analysis of the system. Specifically, we use the Systems-Theoretic Process Analysis (STPA) \cite{leveson2011engineering} hazard analysis technique to identify the potential unsafe scenarios in an AV with ACC and LKAS mechanisms. The identified unsafe scenarios are then translated into fault injection campaigns where the potentially unsafe system contexts (defined by the unique combinations of state variables) are used as conditions under which the faults are injected into the control software. We compare the performance of the STPA guided fault injection approach in covering hazardous scenarios to a random fault injection approach. % A set of unsafe scenarios with faulty effects have been designed (based on some potential hazards) and applied to the openpilot, and the resilience of its control mechanism has been evaluated. 

%One of the major challenges It indicates that significant amount of testing is required for an AV before final deployment. Current practice in testing and safety validation of AVs involves road testing. But, the safety-critical faults causing hazards and incidents are rare and it might take forever for an AV to experience them during road test experiments. Moreover, with increasing complexity of software in AVs, there are still several challenges in testing and safety validation. ... % 

%\todo[inline]{The safety-critical faults causing hazards and incidents are rare and it might take forever for an AV to experience them during road test experiments} 
%\todo[inline]{look at the proposal introduction and the motivation for using STAMP and re-write this part!--DONE}

%Being one of the most complex safety-critical cyber physical systems, AV depends on the correct and safe execution of the software and hardware that are subject to naturally occurred faults. 

%In this paper, the performance of a self-driving software has been studied and evaluated under different faulty scenarios injected to the sensor data.

%- What have been done? 
%\todo[author=HA,inline]{Distracting, Maybe move to the related work section}

In summary, this paper makes the following contributions: 
\begin{itemize}
    \item STPA based hazard analysis technique is applied to identify the unsafe control actions that may cause safety hazards in ACC and LKAS of AVs.
    \item An open-source testbed for safety validation of AVs is developed by integrating the \textit{openpilot} PC simulator with a computer vision based lane marker detection module and a software fault injection framework that mimics the effect of faults on the RADAR and car sensors as well as real-world environmental conditions impacting camera input, such as rain, fog, snow, occlusion, and blur.
    \item A strategic software fault injection framework based on STPA is developed where the locations and triggers for injecting the faults are driven by the unsafe control actions and critical system context identified during the hazard analysis process. We show that using this technique, the probability of generation of unsafe scenarios and hazard coverage is increased and, thus, the fault injection space is reduced. 
    \item Resilience of the open-source self-driving agent \textit{openpilot} against faults in the RADAR, camera, and car sensors is evaluated by characterizing the propagation of the faults into the system and assessing the ability of the control and safety mechanisms in masking and detection of unsafe scenarios and timely alert on possible safety hazards. 
    %\item A vision based lane marker detection algorithm has been implemented and integrated with PC simulator of the openpilot to be used as the vision module.
    %\item Synthesized road images to mimic real world scenarios like rain, fog, snow, occlusion, contrast, brightness, and blur. The performance of the openpilot has been evaluated under these environmental irregularities.
\end{itemize}
\vspace{-0.5em}
\section{System Overview}
\begin{figure*}[!ht]
  \centering \includegraphics[scale=1.0]{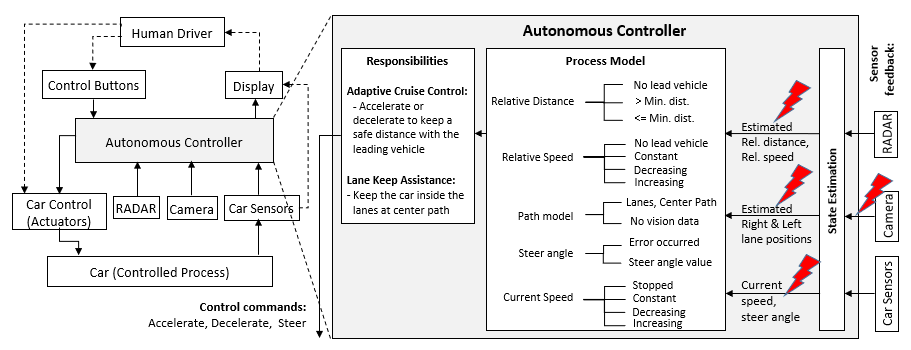}
  \caption{High-level control structure of an AV with ACC and LKAS mechanisms: The autonomous controller is expanded to illustrate the control algorithm responsibilities and process model (state variables) used in the \textit{openpilot} software. The targeted locations for fault injection are highlighted.}
  \label{fig:op_overall}
  \vspace{-2em}
\end{figure*}

\textit{Openpilot} is an open-source alpha quality driving agent introduced by Comma.ai \cite{comma.ai} for research purposes. It is designed with both ACC and LKAS capabilities, and using an additional hardware EON Dashcam DevKit \cite{openpilot:HW} can control the gas, brake, and steering on certain car models, including Honda Civic 2016-2018, Acura ILX 2016, and Toyota RAV4 2016. The left part of Fig. \ref{fig:op_overall} shows the overall system architecture of an AV with ACC and LKAS features. The interactions among the human driver, the autonomous controller, and the controlled process (vehicle) are depicted. The functional control diagram of the autonomous controller is further expanded in the right part of Fig. \ref{fig:op_overall}. In this figure, the interactions of the human driver with the autonomous controller and the vehicle (when autonomous control is offline) are shown with dashed lines with arrows. These interactions will not be studied in this work, as our main focus is to evaluate the resilience of the autonomous controller to the faults and environmental conditions affecting the sensor data. %It also shows the locations targeted for fault injection. 
\par
\textbf{Control Mechanisms:} \textit{Openpilot} uses vehicle's radar to estimate the lead vehicle's position and speed to maintain a safe distance with it, while driving at desired cruise speed. For the lane keeping assistance, it uses phone camera to capture the sequences of images of the road and detect the lane markers. The control software consists of several threads, including \textit{controls} thread, \textit{radar} thread, \textit{vision} thread, and \textit{sensor} thread. \textit{Controls} thread is the main control loop (running at 100Hz) that communicates with the lower level car control mechanisms. \textit{Radar} thread parses the messages from the car radar and derives the positions and velocities of up to two lead vehicles. The ACC algorithm uses this information to calculate the target acceleration or deceleration. It also uses the information on the current speed of the host vehicle from the \textit{sensor} thread. In the recent versions of \textit{openpilot}, the relative distance of the lead vehicle is also captured based on the camera data. The autonomous controller uses the relative distance information coming from both the radar and vision modules, but the ACC still mainly relies on the radar data for estimating relative distance. \textit{Vision} thread runs a neural network based algorithm to detect the road lane markers and outputs the best estimate of the path. LKAS lateral control algorithm uses this path information and current steer angle information (collected from the \textit{sensor} thread) to derive the required steer torque to be applied to keep the vehicle in the center of the lane. \par
%Besides the basic functionality of the ACC and LKAS, we have also studied 

\textbf{Safety Mechanisms:} The existing safety mechanisms of \textit{openpilot} include an alert manager module responsible for generating warning messages to alert the driver in case of any emergency situations. The majority of these warnings are related to the technical issues with the car, such as brake unavailable, gas unavailable, or door open. In this work, we mainly focus on the alerts related to the functionality of the ACC and LKAS as well as the radar and vision modules, e.g., CAN errors, communication errors, steer saturation errors. \textit{Openpilot} software has also been recently equipped with a forward collision warning (FCW) system that triggers a warning if the vehicle needs to decelerate quickly to avoid a rear-end collision with the lead vehicle. %\cite{openpilot:FCW}. 
In this study, we evaluate the functionality of ACC and LKAS systems and the performance of the FCW system in case of faulty inputs.% to the radar, car sensors, and the vision module.
\par

\textbf{PC Simulator:}  \textit{Openpilot} has a PC simulator that enables running its control software without the actual car and radar sensors in the loop. The simulator mimics the functionality of the actual car and radar module, but instead of modeling the camera functionality and including vision processing, it uses a simplified model of an infinitely long straight lane with the vehicle always being at the center of the lane (the actual vision module used in \textit{openpilot} \cite{DBLP:journals/corr/SantanaH16} is not open-source). In order to simulate more realistic road conditions, we developed a simulated vision module and integrated it with the \textit{openpilot} PC simulator.

\textbf{Vision Module:} %As mentioned in previous section, the PC simulator of the openpilot contains a sample car model and a radar model. But instead of a vision algorithm, it uses a long straight lane. So the trajectory of the vehicle will always be a straight line at the center of the lane\cmmnt{(Fig. \ref{fig:iolkasb})}, which is not realistic. So, we have implemented a lane marker detection algorithm and used it as the vision model for this simulator. 
Our simulated vision module takes RGB road images (video frames) as input and generates the positions of the left and right lane markers on the road which is used by the \textit{openpilot} LKAS system. The lane marker detection algorithm exploits histogram of oriented gradient (HOG) \cite{Vision:HOG} and neural networks (NN) \cite{bishop1995neural} to process the pre-collected images of the road. %Now, the trajectory will change depending on the input images\cmmnt{ (Fig. \ref{fig:iolkasa})}. 
In our experiments, we used the video frames from the Caltech lanes dataset \cite{lane:dataset}, but any other dataset is applicable here. \par
%For the vision model, a HOG-NN based algorithm has been used to detect lane markers from the road images. 

\begin{figure}[!htb]
    \centering
    \begin{subfigure}[b]{0.2\textwidth}
        \centering
        \includegraphics[width=\textwidth]{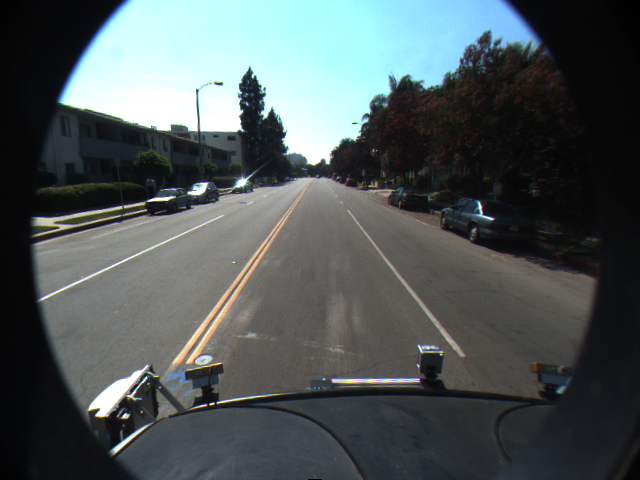}
        \caption{}
        \label{fig:im_orig}
        \vspace{-1.2em}
    \end{subfigure}%
    \hfill%
    \begin{subfigure}[b]{0.2\textwidth}
        \centering
        \includegraphics[width=\textwidth]{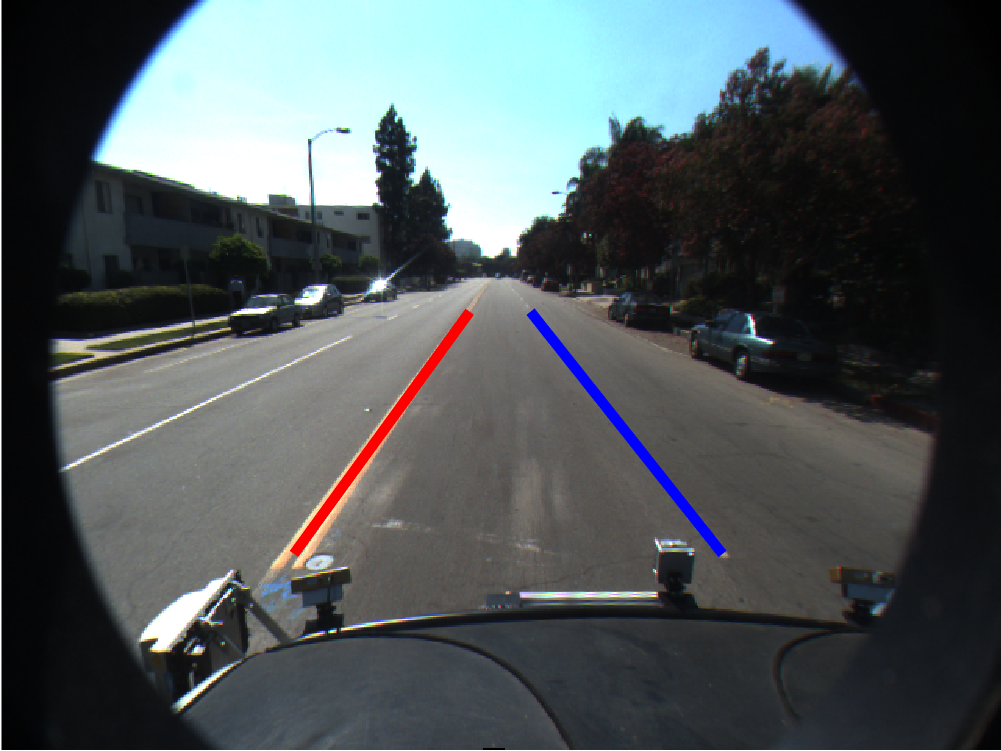}
        \caption{}
        \label{fig:im_res}
        \vspace{-1.2em}
    \end{subfigure}
    \caption{A sample output from the lane marker detection algorithm, (a) input video frame, (b) detected left (red) and right (blue) lane markers.}
    \label{fig:vision_result}
\end{figure}

HOG was introduced in \cite{Vision:HOG} for human detection in images, but its application can be extended to other object detection and image segmentation techniques. The feature extraction procedure using HOG starts with a sliding detection window which is divided into equally spaced blocks. From each block, gradient vectors in both horizontal and vertical directions are calculated and put in a histogram which is used as the gradient feature. ${L2}$ normalization method is used for feature normalization to make it invariant to the illumination changes. A simple NN (${patternnet}$ in MATLAB) with an input layer, one hidden layer with 10 nodes (MATLAB default) and an output layer was trained using these features. The NN classifies the detection windows from an input image into two classes, (i) with lane marker, and (ii) without lane marker. The output from the HOG-NN are the candidate image blocks that are part of the lane markers. Then to extract the lane markers from each block, Sobel edge detector with horizontal derivative (to detect vertical edges) \cite{Sobel:HGrad} is used, which gives a binary image with white pixels representing the edges of the lane markers. To separate the left and right lane markers, a density-based clustering algorithm DBSCAN \cite{Ester:DBSCAN} is used. Assuming the vertical line drawn through the middle of the image as the current path of the vehicle, the immediate left and right lines from the middle line can be considered as the left and right lane markers. The detected lane markers in a sequence of frames constitute the path model that is used by the \textit{openpilot} LKAS system. Fig. \ref{fig:vision_result} shows a sample output from the lane marker detection algorithm. 

\begin{figure}[!ht]
  \centering \includegraphics[scale=0.58]{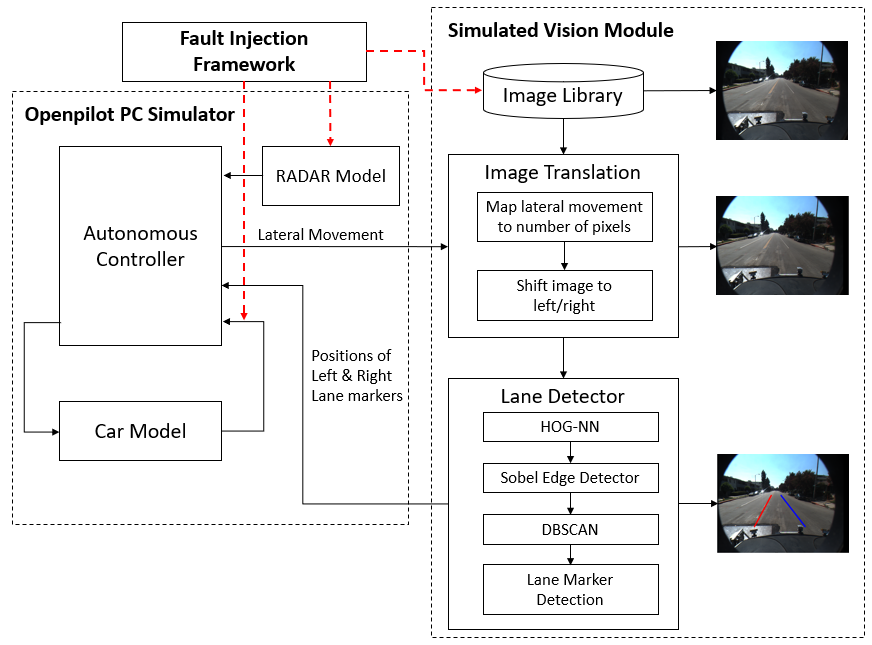}
  \caption{Extended \textit{openpilot} PC simulator \cite{openpilotVis:Git} with our simulated vision module (lane marker detection algorithm) and fault injection framework.}
  \label{fig:vis_op}
  \vspace{-1.8em}
\end{figure}

Fig. \ref{fig:vis_op} shows an overview of the extended \textit{openpilot} simulator with the simulated vision module and fault injection framework. %It also depicts the interaction between the openpilot's autonomous controller and the vision module. 
The Lane Detector block implements the lane marker detection algorithm described above. The Image Library contains the dataset of images used for simulation experiments. The Image Translation block is used to mimic the effects of changing steering angle on the view of camera and horizontally translates the input images based on the lateral movement of the simulated vehicle.

\begin{comment}
\begin{figure*}[!ht]
    \centering
    \begin{subfigure}[b]{0.3\textwidth}
        \centering
        \includegraphics[width=\textwidth]{Images/plotsRadarIO}
        \caption{}
        \label{fig:iorad}
    \end{subfigure}%
    %\hfill%
    \begin{subfigure}[b]{0.3\textwidth}
        \centering
        \includegraphics[width=\textwidth]{Images/plotsLkasIOb}
        \caption{}
        \label{fig:iolkasb}
    \end{subfigure}
    \begin{subfigure}[b]{0.3\textwidth}
        \centering
        \includegraphics[width=\textwidth]{Images/plotsLkasIOa}
        \caption{}
        \label{fig:iolkasa}
    \end{subfigure}
    \caption{Inputs and outputs of (a) ACC, (b) LKAS before adding vision model, and (c) LKAS after adding vision model}
    \label{fig:plotsIO}
\end{figure*}
\end{comment}

\section{Methodology} \label{methodology}
This section describes the steps taken for system hazard analysis using STPA and design of an STPA guided fault injection framework for safety validation.
\vspace{-0.5em}
\subsection{System-theoretic Hazard Analysis}
We used STPA to identify the potential safety hazards of the ACC and LKAS enabled AVs and the causal factors leading to such hazards. The main task of the ACC is to maintain a safe distance with the lead vehicle while moving at cruise speed, and the LKAS keeps the vehicle inside the lanes. Based on the functionality of ACC and LKAS, we first classify the AV accidents into three types:
\vspace{-0.5em}
\begin{itemize}
\item \textbf{A1:} rear-end collision with the lead vehicle,
\item \textbf{A2:} causing traffic congestion or collision with the trailing vehicle, and 
\item \textbf{A3:} side collision with other vehicles or road-side objects. 
\end{itemize}

Three types of system hazards or set of system states could result in those accidents:
\begin{itemize}
    \item \textbf{H1:} AV violates maintaining safety distance with the lead vehicle that may result in A1.
    \item \textbf{H2:} AV decelerates to a full stop although there is no lead vehicle which may lead to A2.
    \item \textbf{H3:} AV goes out of lane which may lead to A3.
\end{itemize}

\begin{figure*}[!htb]
    \centering
    \begin{subfigure}[b]{0.3\textwidth}
        \centering
        \includegraphics[width=\textwidth]{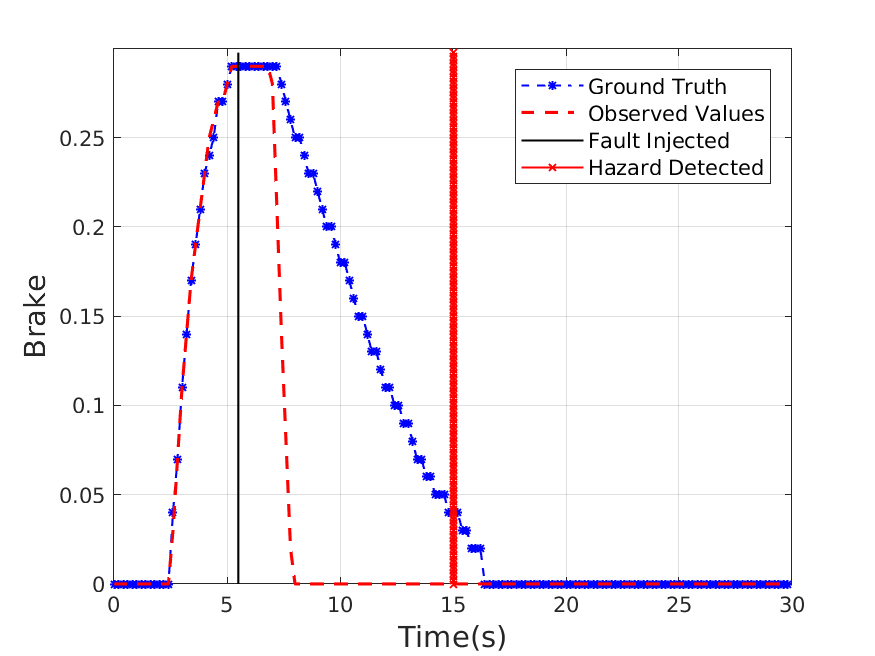}
        \caption{}
        \label{fig:UCAbrake}
        \vspace{-1em}
    \end{subfigure}%
    %\hfill%
    \begin{subfigure}[b]{0.3\textwidth}
        \centering
        \includegraphics[width=\textwidth]{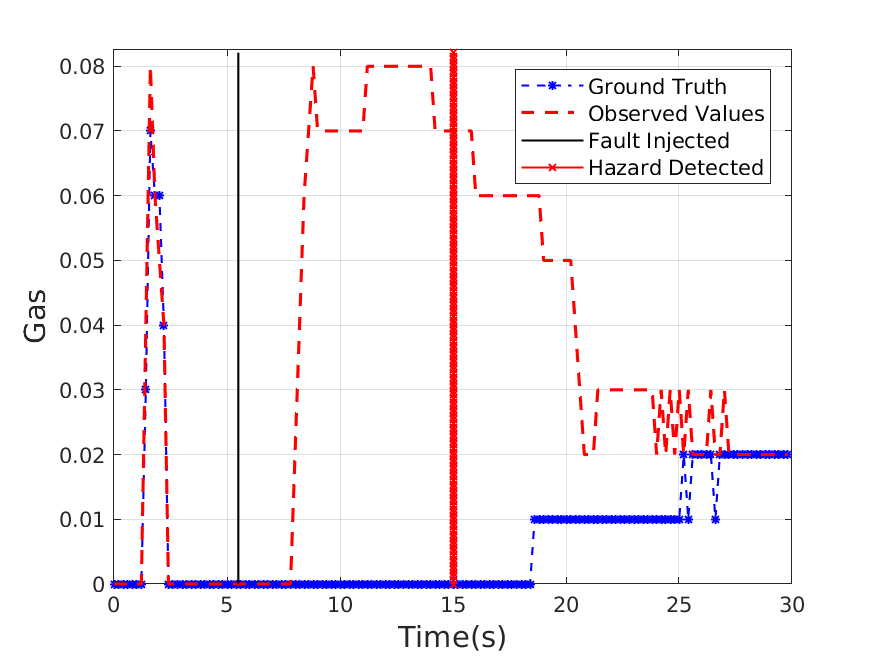}
        \caption{}
        \label{fig:UCAgas}
        \vspace{-1em}
    \end{subfigure}
    %\hfill%
    \begin{subfigure}[b]{0.3\textwidth}
        \centering
        \includegraphics[width=\textwidth]{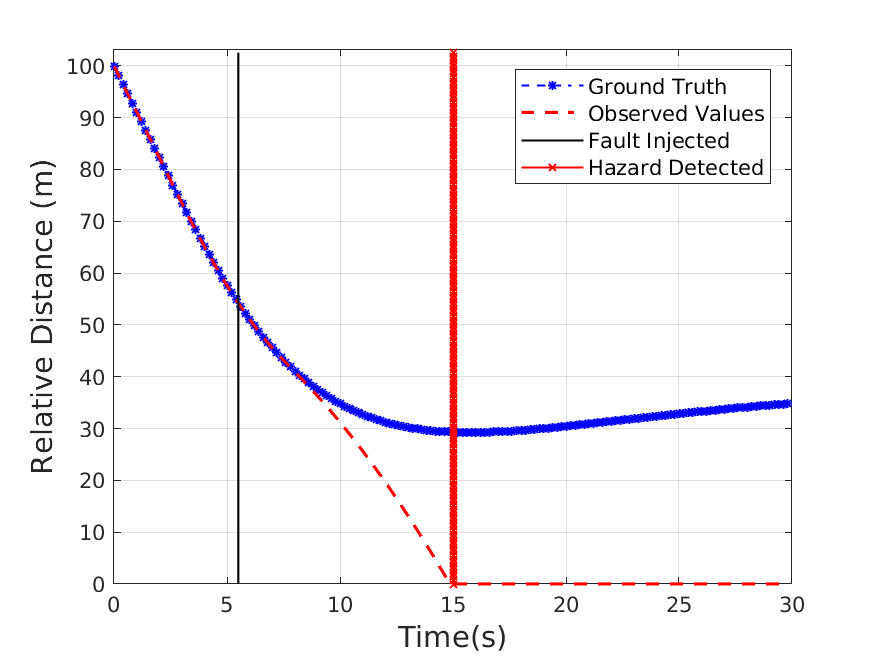}
        \caption{}
        \label{fig:hazard}
        \vspace{-1em}
    \end{subfigure}
    \caption{Example of unsafe control actions (a) Brake provided, but for a short period, (b) Gas provided, but at a wrong time, leading to a hazard, (c) Collision with the lead vehicle (relative distance becomes zero).}
    \label{fig:UCA}
    \vspace{-2em}
\end{figure*}

Next step is to model the overall hierarchical safety control structure of the \textit{openpilot} system, as shown in Fig. \ref{fig:op_overall}. We identified unsafe control actions that lead to hazardous scenarios and potential causal factors for the unsafe actions by examining inputs and outputs across different loops of the control structure. Outputs of the ACC and LKAS are gas/brake and steer-torque, respectively. Following the STPA categories, the potential unsafe control actions can be classified as, "Acceleration", "Deceleration", "Steer" commands (using gas/brake/steer-torque): (i) required but not provided, (ii) not required but provided, (iii) provided but with incorrect timing, and (iv) provided for a wrong duration \cite{leveson2011engineering}. For example, Fig. \ref{fig:UCA} shows the simulation of a scenario where faulty radar data caused unsafe control actions - brake ("Deceleration") provided for short period (Fig. \ref{fig:UCAbrake}) and gas ("Acceleration") provided at a wrong time (Fig. \ref{fig:UCAgas}), which further led the host vehicle collide with the lead vehicle (A1/H1) (Fig. \ref{fig:hazard}).\par %Here, incorrect radar data can be considered as a potential causal factor for these hazards.

\subsection{STPA Guided Fault Injection}
Because of the complexity of the AVs control software, the test space for fault injection testing is often huge. For example, \textit{openpilot} software used in this work consists of 146 python files with over 22,000 lines of code. Thus, it might take a long time for a random fault injection approach to generate hazard scenarios of interest. To address this challenge, we adapt a strategic fault injection approach where the target locations and triggers (or conditions) for injecting the faults are driven by the hazard analysis of the system. 

Our fault injection framework evaluates the resilience of AV controller by simulating the unsafe scenarios (unsafe control actions and potential causal factors) identified by STPA using software-implemented fault injection. In this paper, we mainly focus on the causal factors related to the faulty inputs which might directly affect AV controller perception functionality. This is motivated by a recent study \cite{subho2018hands} of the field data on AVs from the California Department of Motor Vehicles that found the faults in the perception systems are responsible for the majority of disengagements per mile across manufacturers (44\% of reported disengagements). As shown in Fig. \ref{fig:op_overall}, three main sources of the inputs to the \textit{openpilot} controller are radar, camera, and car sensors. The controller uses the information provided by these sensors to calculate the required gas/brake and steer torque. So, our target modules for injecting faults are (i) Vision module, (ii) RADAR module, and (iii) Car sensors. In the case of vision module, faults are injected to both input images and output (position of the lane markers). For radar and car sensors, only the outputs produced (sent to controller) by these modules are subject to fault injection. \par

Using STPA the context or set of system states under which the unsafe control actions occur can be identified from the combinations of the state variables used to describe the process model in the system control structure. We use the identified context for unsafe control actions as the triggers for fault injection. Fig. \ref{fig:op_overall} shows that the state variables of the autonomous controller process model are relative distance and speed, current speed of the host vehicle, steer angle, and path model. For deducing the system context we use a subset of these variables, including relative distance, relative speed, and current speed. Relative speed is represented by $HV_{speed} - LV_{speed}$ in which $HV_{speed}$ and $LV_{speed}$ represent the speed of the host vehicle and lead vehicle respectively. Relative distance and host vehicle's speed can be combined into headway time ($Relative Distance/Current Speed$). Headway time (HWT) is the time gap between the host vehicle and lead vehicle. Table \ref{tab:context} shows an example context table for the "Accelerate" and "Decelerate" control actions. %The fault injection triggers are designed based on the context deduced from different combination of HWT and relative speed. 
The "Context" column in each row of the table represents the condition (using combinations of HWT and relative speed variables) that is used as the trigger for injecting the faults. For example, to generate an unsafe "Accelerate" command which might lead to hazard H1 (row 2), faults are injected to the input images when the relative speed \textit{(RS)} of the vehicle is "high" and the \textit{HWT} is "less than a safe distance (\textit{safeHWT})".

\begin{table}
    \renewcommand{\arraystretch}{1.3}
    \centering
    \caption{Example STPA context table for \textit{Accelerate} or \textit{Decelerate} control actions
    }
    \begin{tabular}{|m{4em}|c|m{5em}|m{4em}|m{4em}|}
        \hline
         \multirow{2}{4em}{\textbf{Control Actions}} & \multicolumn{2}{|c|}{\textbf{Context}} & \multicolumn{2}{|c|}{\textbf{Hazardous Control?}} \\
         \cline{2-5}
         & Headway Time (HWT\footnotemark[1]) & Relative Speed (RS\footnotemark[2]) & Command Not Provided & Command Provided  \\
         \hline
         \multirow{4}{*}{\textbf{Accelerate}} & \multirow{2}{*}{$HWT \leq safeHWT$\footnotemark[3]} & $RS \leq 0$ & No & No \\
         \cline{3-5}
         & & $RS > 0$ & No & \textbf{Yes-H1} \\
         \cline{2-5}
         & \multirow{2}{*}{$HWT > safeHWT$} & $RS \leq 0$ & \textbf{Yes-H2} & No \\
         \cline{3-5}
         & & $RS > 0$ & No & \textbf{Yes-H1} \\
         \hline
         \multirow{4}{*}{\textbf{Decelerate}} & \multirow{2}{*}{$HWT \leq safeHWT$} & $RS \leq 0$ & No & No \\
         \cline{3-5}
         & & $RS > 0$ & \textbf{Yes-H1} & No \\
         \cline{2-5}
         & \multirow{2}{*}{$HWT > safeHWT$} & $RS \leq 0$ & No & \textbf{Yes-H2} \\
         \cline{3-5}
         & & $RS > 0$ & \textbf{Yes-H1} & No \\
         \hline
    \end{tabular}
    \label{tab:context}
    \vspace{-2.5em}
\end{table}

\footnotetext[1]{$HWT=Relative Distance/Current Speed$}
\footnotetext[2]{$RS=Current Speed-Lead Vehicle Speed$}
\footnotetext[3]{$safeHWT \approx 2.0s\sim 3.0 s$}
\vspace{-0.5em}
\subsection{Fault Injection Framework}\label{fault_injection_framework}
As discussed before, %in this work the input modules (shown in Fig. \ref{fig:op_overall} using red symbols) are the target of the fault injection. 
we simulate the effect of faults and environmental conditions by injecting faults into the inputs and outputs of the vision module, the RADAR, and car sensor. To do this, we have generated a library of faulty images based on the effects described in  Table \ref{table:im_effects}. During simulation, the vision module reads the images with different faults based on the specified fault model. For the RADAR, car sensors, vision outputs, faults are injected using a compile-time fault injection approach similar to \cite{Alemzadeh:Safecomp}. First, a fault library is manually compiled based on the fault models (defining the locations and values for injection) and the STPA context table (defining the injection triggers). Then the fault injection campaigns are automatically generated by creating faulty codes to be added to the target locations within the software.

On each run of the experiments, the injector adds the faulty codes to the specified location and then executes the \textit{openpilot} simulator. 
%The HWT between the lead vehicle and the host vehicle is  Rows in the context table are used as the trigger for injecting the faults. 
The time of triggering the fault, the simulation results, and the alerts generated from the \textit{openpilot} are logged for further analysis. The system outputs, including gas/brake (ACC) and steering torque (LKAS) values were recorded for each experiment and compared with the outputs generated from the simulations without faults (ground truth). Any deviation from the ground truth was considered as the fault being manifested. 

Separate checkers are placed within the code to detect and identify the occurrence of the three hazards (H1, H2, H3). The H1 hazard (violation of safety distance) is checked based on the relative distance with respect to the lead vehicle. Speed and relative distance are used to check whether the H2 (sudden stop) hazard has happened or not. Deviation from the center path is recorded to check the out of lane hazard (H3). If any of these three hazards occur, the time of hazard is also recorded. \par

%As the functionality of the target modules are different from each other and the characteristics of the data handled by them are also not similar, we have used different fault models for different modules. 
The fault models and the parameters used for injecting faults into different modules are described next.
\subsubsection{\textbf{Vision Module}}
In the vision module, faults are injected into the input frames as well as the outputs, including the left and right lane positions and the predicted path of the vehicle. 
By injecting faults to input images, we measure the resilience of the ML lane detection algorithm. By injecting faults to the outputs we evaluate the response of the LKAS system if the vision module produces erroneous outputs. To implement these faults, random values are added/subtracted to/from the path model variables. It is also possible that the camera becomes unavailable or the communication between vision data and control thread gets disconnected. To simulate these faults, the vision data is made unavailable after a certain time (after the trigger) of the simulation. \par
Faults injected to the input images of the vision module simulate real-world environmental conditions such as rain, fog, snow, and occlusion created by mud/snow on camera. We also use other transformations like contrast change, brightness change, and adding blur to the images to evaluate the resilience of LKAS. %These faults are added to the road image sequences and these images are fed to the vision algorithm described in the previous section. 
More details on adding these effects to the input images are described below:

\textit{Rain}: To add the effect of rain, rain streaks from the database \cite{RainStreak:Garg} are used.  Rain streaks are added in random positions. For varying the density of the streaks, a 'thickness' parameter is used. The value of 'thickness' varies from 0 to 10, 0 means no rain and 10 indicates maximum density (10,000 streaks per image). The angle of the streaks can also be varied. To adjust with the background, the contrast of the rain streak is reduced and motion blur is added to the streaks. Besides adding the streaks, rain has also other effects (e.g., cloudy, gloomy) on the environment. To add those effects, we apply Gaussian blur and contrast reduction (varies with thickness) to the background of image. The values for the number of streaks, standard deviation ($\sigma$) of Gaussian blur, and contrast gain ($\alpha$) for the background are set empirically.

\textit{Fog}: For creating the foggy effect, a uniformly distributed random noise is added to the images. Also a Gaussian blur is applied to the noise to add the blurriness effect. Density of the added noise varies with 'thickness' parameter. To create the haziness, contrast of the background image is varied.

\textit{Snow}: Small blocks (size 2x2) of white pixels are added to the random locations of the image to simulate the effect of falling snow. Similar to rain and fog, number of snow varies with 'thickness'. Gaussian and motion blur are applied to the white blocks. Snow-fall also creates the haziness like the fog. So, the procedure of adding fog (described previously) is applied on the background images first, and then falling snows are added on top.

\textit{Occlusion}: In real environment, there are many factors that can cause occlusion. In this work, we consider the occlusions created by mud/snow stuck on the camera. To create the effects of occlusions, circular blobs (dark) are added in random locations of the image. Number of the blobs can be varied and the radius of the blob is randomly chosen from the range $[5, 50]$ pixels. To make the blobs blurry, Gaussian blur is applied.

\textit{Other Effects}: Besides the faults described above, we also apply some other transformations from \cite{DeepTest:Tian} to the input images. These are changes in contrast, brightness, and blur (averaging, Gaussian, median).

\par
Table \ref{table:im_effects} summarizes the parameters used to inject the mentioned effects to the input images. The last column shows the faulty images along with the lanes detected by the lane marker detection algorithm.

\begin{table}[!t]
\renewcommand{\arraystretch}{1.3}
\caption{Parameters used to add environmental effects and faults to the image shown in Fig. \ref{fig:im_orig}}
\label{table:im_effects}
%\centering
\setlength\tabcolsep{5pt}
\begin{tabular}{|>{\centering\arraybackslash}m{9mm}|>{\centering\arraybackslash}m{47mm}|>{\flushleft\arraybackslash}m{20mm}|}
\hline
\textbf{Fault} &\textbf{Parameters} & \textbf{Results} \\
\hline
Rain 
&
%Contrast (streak) \newline Contrast (environment) \newline Motion Blur (streak) \newline Gaussian Blur (environment)
\begin{itemize}[leftmargin=*]
    \item Contrast (streak): $\alpha$ = 0.1 $\sim$ 0.2
    \item Contrast (environment): $\alpha$ = 0.65 $\sim$ 0.75
    \item Motion Blur (streak): $l$ = 2 pixels, $\theta$ = 45 degrees
    \item Gaussian Blur (environment): $\sigma$ = 0.5 $\sim$ 1.5
\end{itemize}
&
\includegraphics[width=0.11\textwidth]{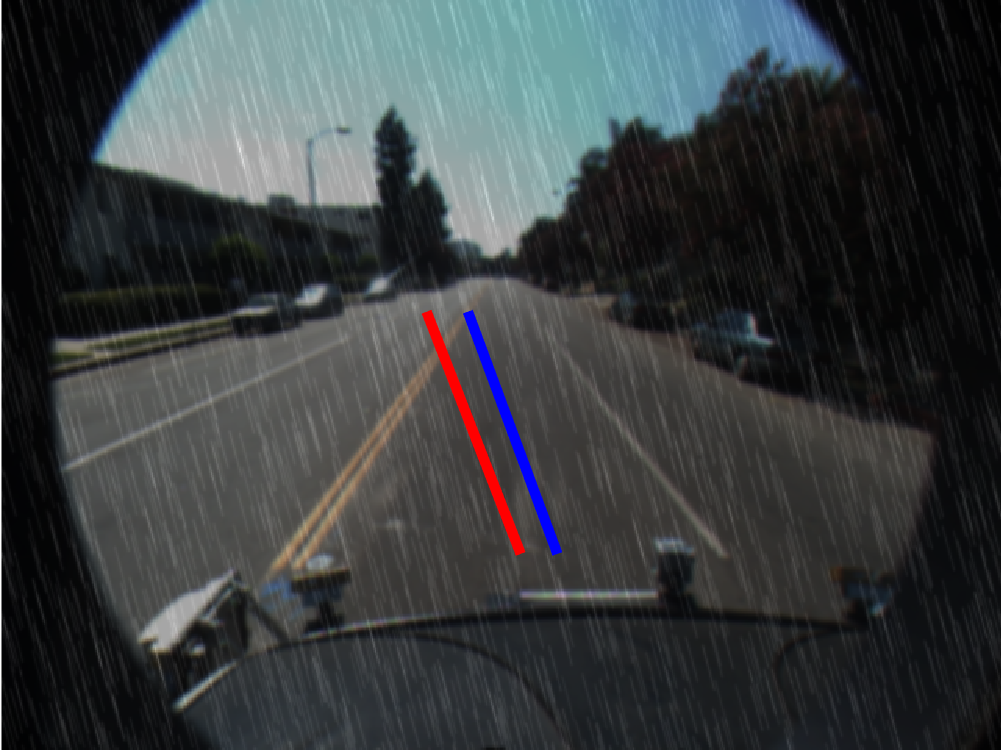} \\
\hline
Fog 
& 
\begin{itemize}[leftmargin=*]
    \item Gaussian Blur (droplets): $\sigma$ = 5 $\sim$ 7
    \item Contrast (environment): $\alpha$ = 0.4 $\sim$ 0.9
\end{itemize}
&
\includegraphics[width=0.11\textwidth]{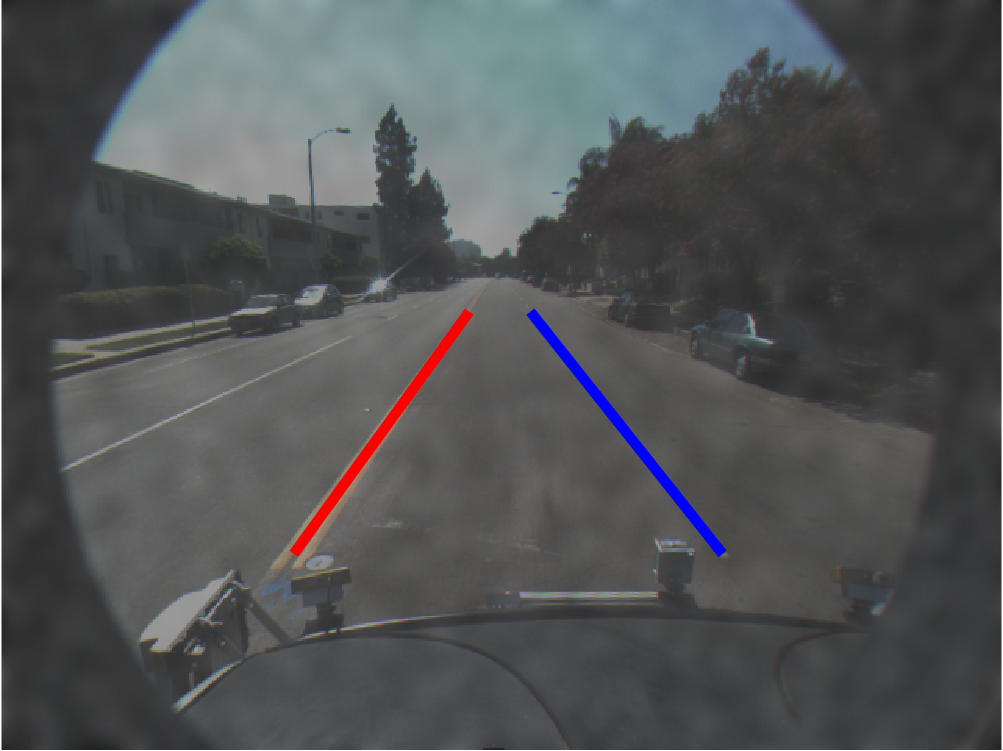} \\
\hline
Snow 
&
\begin{itemize}[leftmargin=*]
    \item Gaussian Blur (snow): $\sigma$ = 0.5 $\sim$ 1.5
    \item Motion Blur (snow): $l$ = 5 pixels, $\theta$ = 75 degrees
    \item Gaussian Blur (Haze): $\sigma$ = 5 $\sim$ 7
    \item Contrast (environment): $\alpha$ = 0.4 $\sim$ 0.9
\end{itemize}
&
\includegraphics[width=0.11\textwidth]{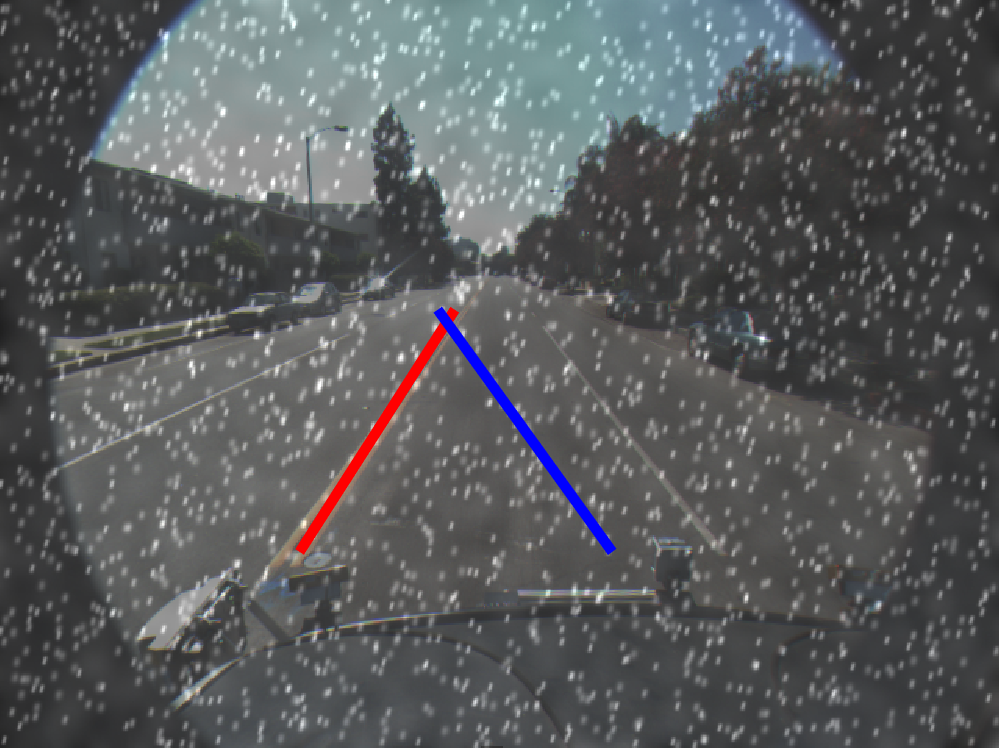} \\
\hline
Occlusion
&
\begin{itemize}[leftmargin=*]
    \item Blobs: Radius = 5 $\sim$ 50 pixels (Random)
    \item Gaussian Blur (blobs): $\sigma$ = 7.0
\end{itemize}
&
\includegraphics[width=0.11\textwidth]{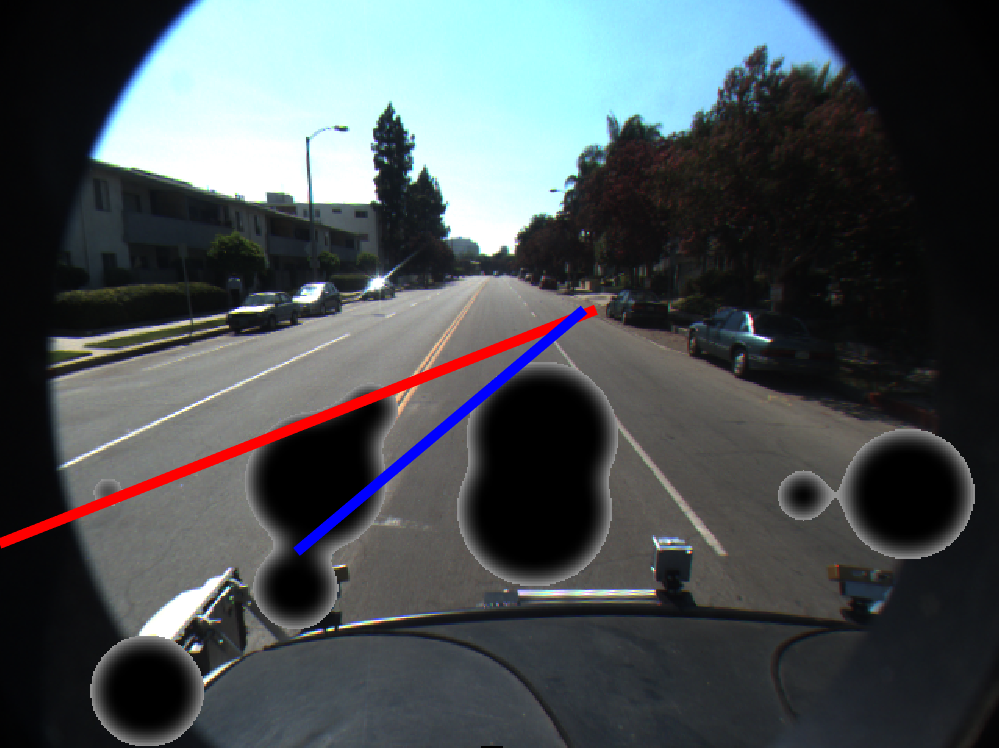} \\
\hline
Contrast
&
\begin{itemize}[leftmargin=*]
    \item Gain, $\alpha$ = 1.2 $\sim$ 3.0
\end{itemize}
&
\includegraphics[width=0.11\textwidth]{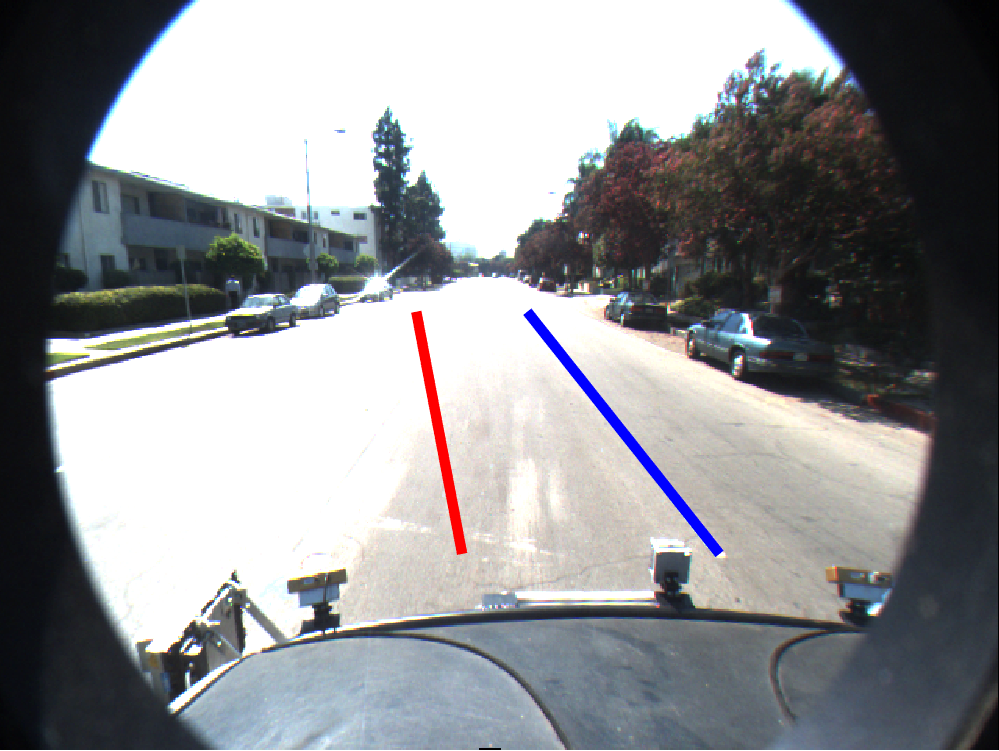} \\
\hline
Brightness
&
\begin{itemize}[leftmargin=*]
    \item Bias, $\alpha$ = 10 $\sim$ 100
\end{itemize}
&
\includegraphics[width=0.11\textwidth]{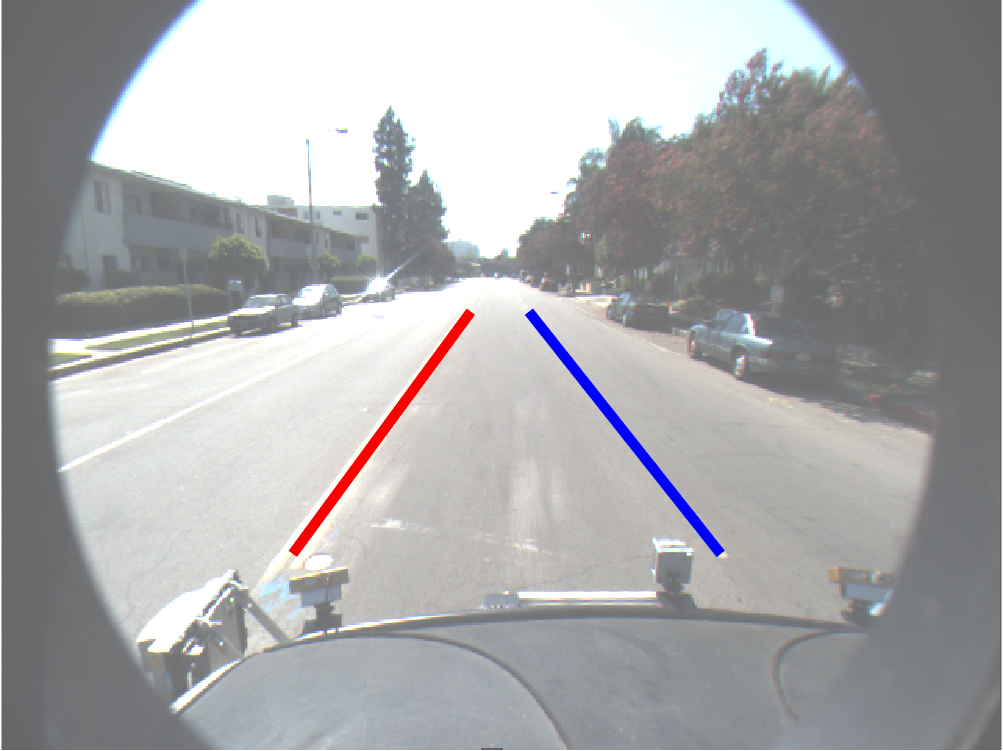} \\
\hline
Blur
&
\begin{itemize}[leftmargin=*]
    \item Averaging: Kernel Size = 3X3, 4X4, 5X5, 6X6
    \item Gaussian: Kernel Size = 3X3, 5X5, 7X7
    \item Median: Aperture Linear Size = 3, 5
\end{itemize}
&
\includegraphics[width=0.11\textwidth]{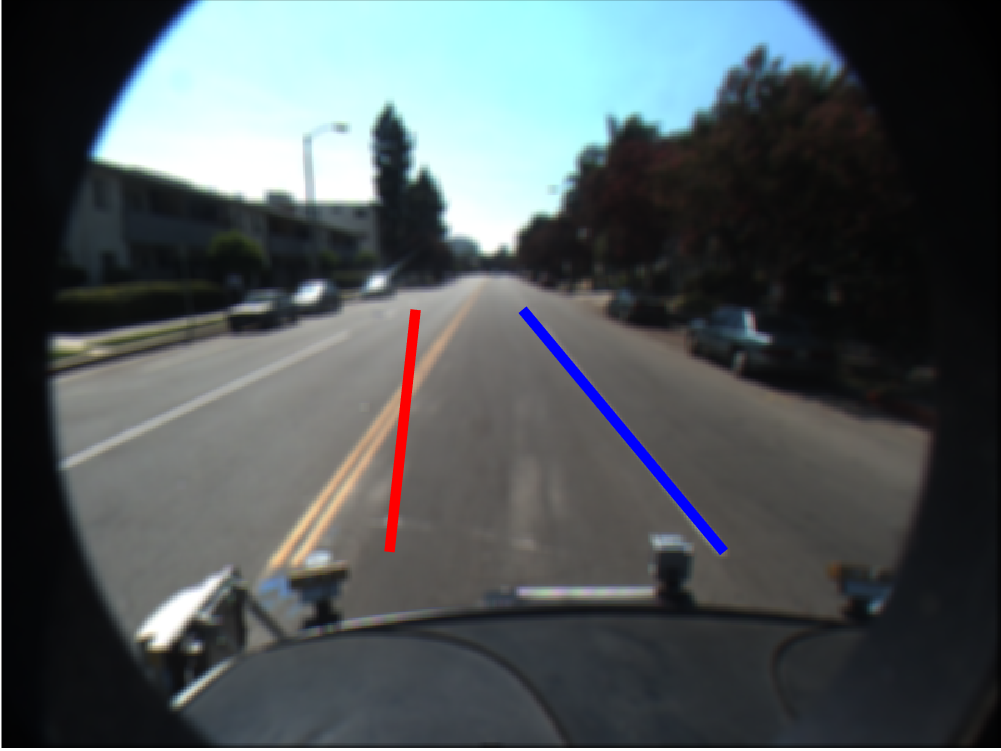} \\
\hline

\end{tabular}
\vspace{-2.1em}
\end{table}

\subsubsection{\textbf{RADAR Module}}
Radar data provides the relative distance and velocity of the lead vehicle that is used by ACC to maintain the safe distance from the lead vehicle. %So, it is important for the radar module to generate accurate information. 
In real-world environments, different types of faults can affect the radar performance. Petit and Shladover \cite{RadarFault:Petit} discussed four types of cyber attacks that can reduce the radar accuracy - chaff, smart material, jamming, and signal repeater. %Although these are mainly attacks to the radar data, 
Similar effects can happen accidentally in real-world environments, causing the radar module to generate erroneous results. For example, effects similar to chaff can be found in rainy or snowy environments. We simulate this effect by adding/subtracting  random values to/from the radar module output. In military applications, object with non-reflective surface has been used to make it invisible to the radar detectors. In the vehicle environments similar phenomenon (invisible vehicle) can happen because of interference among radar signals. Jamming can also occur in the scenarios with high vehicle density, where interference among the radar signals from different vehicles can create the jamming in radar module. This may cause degradation of radar performance and even turning off the radar. We simulate this effect by making the radar data unavailable to the controller. The fourth attack mentioned in \cite{RadarFault:Petit} is a signal repeater which is mainly used to create a ghost object in front of the radar. In real world, rain, snow, dust or dense fog can cause a similar effect \cite{RadarFault:Rasshofer}, i.e., the radar module can detect a ghost vehicle in these kinds of scenarios although there is no actual lead vehicle or other objects on the road.

\subsubsection{\textbf{Car Sensors}}
%RADAR and Vision modules gather data from environment and send the processed data to the controller.
Besides the information from the environment, the AV controller also requires the information of the current status of the vehicle itself. For example, ACC needs the information of current speed of the host vehicle to calculate the target acceleration/deceleration. Or the LKAS needs the current steer angle to learn about the current direction of the vehicle. The AV controller gets this data from the car sensors via CAN messages. In addition to the current speed and steer angle, car sensors provide information about cruise settings, gas/brake pedal status, and internal error flags generated by the car. We only focus on the faults that affect the current speed and steer angle values. These values can be erroneous because of any malfunction in the sensors. We use an additive fault model where random values (positive and negative) are added to the actual speed and steer angle values before sending to the control thread. The offset value added to the actual speed is chosen in such a way that the value of current speed varies between 0mph to twice the desired cruise speed. The steer angle offset value is set empirically between -45\textdegree $\sim$ +45\textdegree. %The ranges of the offset values are set empirically.
\vspace{-0.25em}
\section{Experimental Results}
\subsection{Experimental Setup}
The fault injection experiments were conducted on an x86-64 PC with 16GB RAM running Linux Ubuntu 16.04 LTS. The machine contained an Intel Core i7 CPU @ 3.60GHz. The version v0.4.2 of \textit{openpilot} PC simulator \cite{openpilot:Git} was used. The PC simulator takes around 35 seconds to run 30 seconds of simulated driving. We slowed down the PC simulator to sync it with our simulated vision module which processes the input images at 20 frames per second. On average it took approximately 4.5 minutes to run each experiment. %\todo[inline]{Add the spec for the PC used for running experiments. You can check how we did this in the RAVEN attack paper in DSN 2016} 

We consider different driving scenarios involving a host vehicle with an initial constant speed of 60mph following a lead vehicle with varying acceleration and deceleration behaviors. Specifically, the following five scenarios were simulated in our experiments:
\begin{enumerate}
    \item Lead vehicle is moving with constant speed (40mph).
    \item Lead vehicle is moving with constant low speed (25mph).
    \item Lead vehicle  accelerates and then  slows down.
    \item Lead vehicle  slows down and then  accelerates.
    \item Lead vehicle slows down to a full stop.
\end{enumerate}

In total, 1128 fault injection experiments were conducted for each scenario, leading to a total number of 5640 experiments. The resilience of the target system was evaluated using the following metrics:

\begin{itemize}
    \item \textbf{\textit{Activated Faults}}: If the trigger condition is fulfilled, the fault gets activated and starts affecting the system operation by actually generating erroneous input data.
    \item \textbf{\textit{Manifested Faults}}: If the outputs of the system (e.g., brake, gas, torque) deviate from the expected outputs (ground truth), then the activated fault is considered to be manifested. The manifested faults result in either \textit{Silent Data Corruption (SDC)} or \textit{Hazards}.
    \item \textbf{\textit{Hazard Coverage (C)}}: We measure the coverage of safety-critical faults by the fault injection experiments using hazard coverage, defined as the conditional probability that given activation of a fault in the system, it leads to an unsafe state or hazardous condition.
 %\( C = \text{Prob.}(\text{safety hazard} \mid \text{fault activated})\).Applying basic probability rules, a simplified version of this equation can be achieved: \( C = (\text{Total no. of safety hazards}) / (\text{Total No. of activated faults})\).
    \item \textbf{\textit{Reaction Time}}: Reaction time is the maximum time the driver has for responding to an alert and taking actions and is an important metric for measuring the performance of collision warning systems \cite{Zhang:EvalCAS}. Here the reaction time is defined as the time between raise of an alert by the AV controller and the occurrence of a hazard.
\end{itemize}

We also compared the performance of the STPA guided fault injection vs. random fault injection in terms of fault activation rate and hazard coverage. In the random fault injection, the same number of faults are injected to the same target variables within the control software, but instead of using the STPA contexts as trigger, a random time during simulation is selected as the trigger for injection. Each experiment consisted of injecting the erroneous values to the target variables when the specific trigger is met and continuing the injection for as long as the trigger is active (in case of guided injection) or for the rest of simulation (in case of random injection). %The performance of these two techniques has been compared based on hazard coverage.
\vspace{-0.5em}
\subsection{Outcome Analysis}
 In 3678 out of 5640 (65.2\%) guided injections, faults got activated and actually caused erroneous system inputs. Fault manifestation rate was 99.4\% (3656) with respect to activated faults. Only in around 35.8\% of the activated faults, hazards occurred. The controller generated alerts in 69.5\% of the hazardous scenarios (914 out of 1316), i.e., a large portion (30.5\%) of hazards remained unnoticed. Table \ref{table:summary} summarizes the overall fault injection results. Columns 2-6 contain the separate results for different driving scenarios and column 7 shows the combined results. The remaining columns show the results of random injection experiments. Total number of experiments done for the random injection was the same as the guided technique, however almost all the faults got activated. Hazard coverage for the guided and random fault injection techniques were 35.78\% (1316/3678) vs. 28.97\% (1634/5640), respectively. This implies that using context-based triggers results in relatively better hazard coverage than random-time triggers. However, this hypothesis needs to be further examined with larger number of experiments. Table \ref{table:haz_coverage} shows the comparative hazard coverage achieved by guided vs. random fault injection techniques under different fault models. 

%% the guided experiments of modified context trigger condition
\begin{table*}[!t]
%% increase table row spacing, adjust to taste
\renewcommand{\arraystretch}{1.0}
% if using array.sty, it might be a good idea to tweak the value of
% \extrarowheight as needed to properly center the text within the cells
\caption{Summary of fault injection results for the five driving scenarios (The percentage of activated faults is with respect to the number of injected faults. The other percentages are with respect to the number of activated faults).}
\label{table:summary}
\centering
%% Some packages, such as MDW tools, offer better commands for making tables
%% than the plain LaTeX2e tabular which is used here.
\setlength\tabcolsep{5.5pt}
\begin{tabular}{|p{0.15\linewidth}|c|c|c|c|c|c||c|c|c|c|c|c|}
\hline
\multirow{2}{*}{\textbf{No. / Scenarios}} & \multicolumn{6}{|c||}{\textbf{STPA Guided}} & \multicolumn{6}{|c|}{\textbf{Random}} \\
\cline{2-13}
&\textbf{S1} & \textbf{S2} & \textbf{S3} & \textbf{S4} & \textbf{S5} & \textbf{Total} &\textbf{S1} & \textbf{S2} & \textbf{S3} & \textbf{S4} & \textbf{S5} & \textbf{Total} \\
\hline
\textbf{Injected Faults} & 1128 & 1128 & 1128 & 1128 & 1128 & 5640 & 1128 & 1128 & 1128 & 1128 & 1128 & 5640\\
\hline
 \textbf{Activated Faults} & \makecell{613 \\ (54.3\%)} & \makecell{797 \\ (70.7\%)} & \makecell{867 \\ (76.9\%)} & \makecell{798 \\ (70.7\%)} & \makecell{603 \\ (53.5\%)} & \makecell{3678 \\ (65.2\%)} & \makecell{1128 \\ (100.0\%)} & \makecell{1128 \\ (100.0\%)} & \makecell{1128 \\ (100.0\%)} & \makecell{1128 \\ (100.0\%)} & \makecell{1128 \\ (100.0\%)} & \makecell{5640 \\ (100.0\%)}\\
\hline
 \textbf{Manifested Faults} & \makecell{610 \\ (99.5\%)} & \makecell{795 \\ (99.7\%)} & \makecell{858 \\ (99.0\%)} & \makecell{796 \\ (99.7\%)} & \makecell{597 \\ (99.0\%)} & \makecell{3656 \\ (99.4\%)} & \makecell{1111 \\ (98.5\%)} & \makecell{1070 \\ (94.9\%)} & \makecell{1045 \\ (92.6\%)} & \makecell{1021 \\ (90.5\%)} & \makecell{1028 \\ (91.1\%)} & \makecell{5275 \\ (93.5\%)}\\
\hline
\textbf{Hazards} & \makecell{215 \\ (35.1\%)} & \makecell{291 \\ (36.5\%)} & \makecell{274 \\ (31.6\%)} & \makecell{264 \\ (33.1\%)} & \makecell{272 \\ (45.1\%)} & \makecell{1316 \\ (35.8\%)} & \makecell{314 \\ (27.8\%)} & \makecell{318 \\ (28.2\%)} & \makecell{315 \\ (27.9\%)} & \makecell{256 \\ (22.7\%)} & \makecell{431 \\ (38.2\%)} & \makecell{1634 \\ (29.0\%)}\\
\hline
\textbf{Alerts} & \makecell{236 \\ (38.5\%)} & \makecell{333 \\ (41.8\%)} & \makecell{399 \\ (46.0\%)} & \makecell{336 \\ (42.1\%)} & \makecell{266 \\ (44.1\%)} & \makecell{1570 \\ (42.7\%)} & \makecell{399 \\ (35.4\%)} & \makecell{401 \\ (35.5\%)} & \makecell{400 \\ (35.5\%)} & \makecell{404 \\ (35.8\%)} & \makecell{157 \\ (13.9\%)} & \makecell{1761 \\ (31.2\%)}\\
\hline
 \textbf{Hazards with no Alerts} & \makecell{79 \\ (12.9\%)} & \makecell{92 \\ (11.5\%)} & \makecell{64 \\ (7.4\%)} & \makecell{73 \\ (9.1\%)} & \makecell{94 \\ (15.6\%)} & \makecell{402 \\ (10.9\%)} & \makecell{147 \\ (13.0\%)} & \makecell{166 \\ (14.7\%)} & \makecell{145 \\ (12.9\%)} & \makecell{96 \\ (8.5\%)} & \makecell{323 \\ (28.6\%)} & \makecell{877 \\ (15.5\%)}\\
\hline
 \textbf{Alerts with no Hazards} & \makecell{100 \\ (16.3\%)} & \makecell{134 \\ (16.8\%)} & \makecell{189 \\ (21.8\%)} & \makecell{145 \\ (18.2\%)} & \makecell{88 \\ (14.6\%)} & \makecell{656 \\ (17.8\%)} & \makecell{232 \\ (20.6\%)} & \makecell{249 \\ (22.1\%)} & \makecell{230 \\ (20.4\%)} & \makecell{244 \\ (21.6\%)} & \makecell{49 \\ (4.3\%)} & \makecell{1004 \\ (17.8\%)}\\
\hline

\end{tabular}
\vspace{-2em}
\end{table*}

\subsubsection{Hazard Analysis}
\begin{figure}[!t]
\centering
\includegraphics[width=0.45\textwidth]{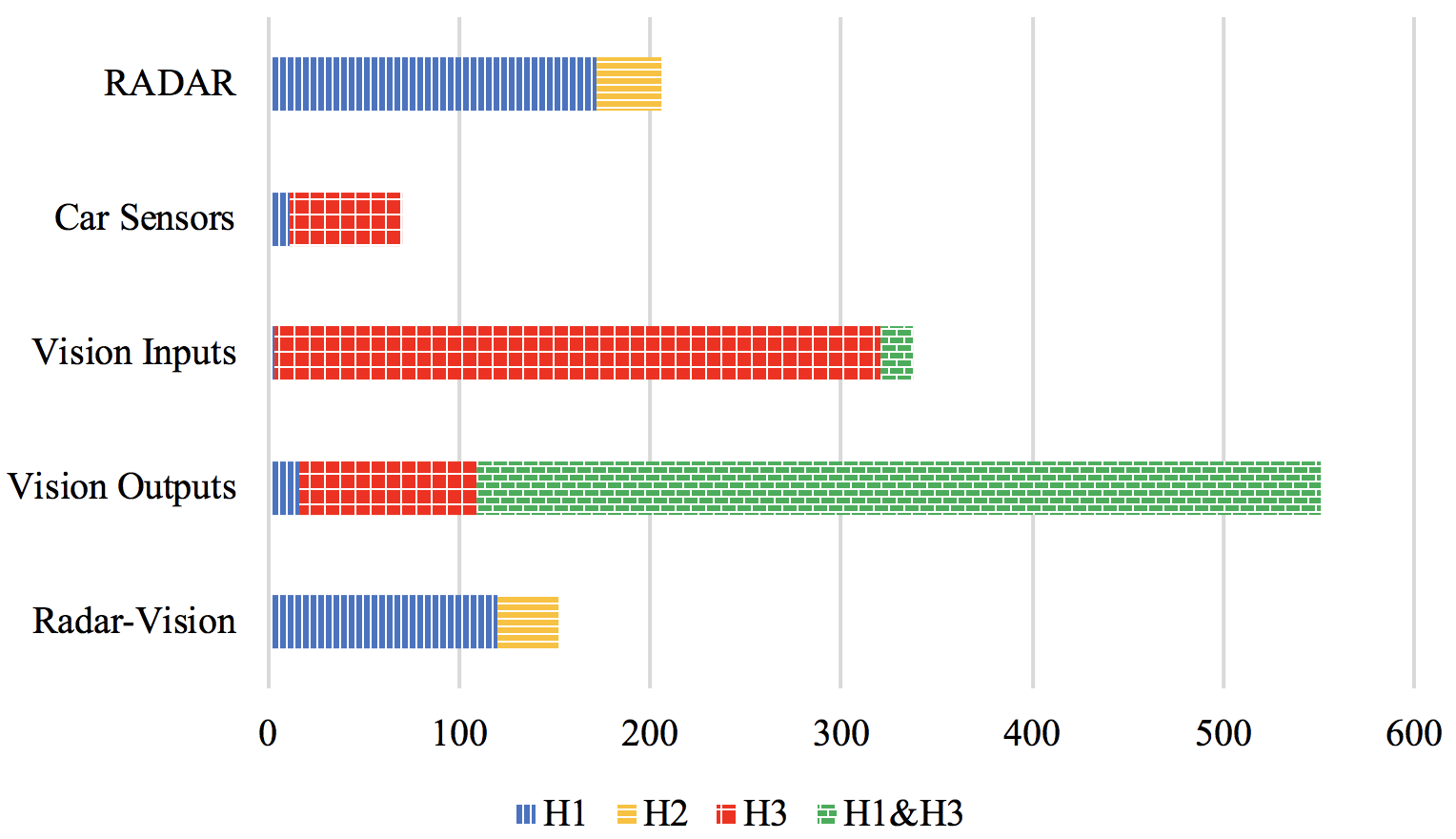}
\caption{Distribution of different fault types across the three hazard scenarios.}
\label{fig:haz_dis}
\vspace{-1.8em}
\end{figure}

As mentioned before, three possible hazards were taken into consideration: violation of safety distance (H1), unnecessary deceleration (H2), and vehicle going out of the lanes (H3). %All three hazards have been checked in each test run. 
The results show that in the majority of cases, a single hazard happened (H1/H2/H3). But in some cases, both H1 and H3 happened in the same experiment. The distribution of the hazards for different fault types are illustrated in Fig. \ref{fig:haz_dis}. H1 and H2 hazards are mainly related to ACC functionality. Thus, the majority of these hazards are caused by the faults injected to the radar module. H3 hazards are caused by the faults in vision module and car (steer) sensors. In a significant number of cases, H1 and H3 occurred together. In those cases, first H3 occurred due to faults in the vision module, then collision with the lead vehicle occurred while the vehicle tried to get inside the lane.

\subsubsection{Time Analysis}
\begin{figure}[!t]
    \centering
    \includegraphics[width=0.45\textwidth]{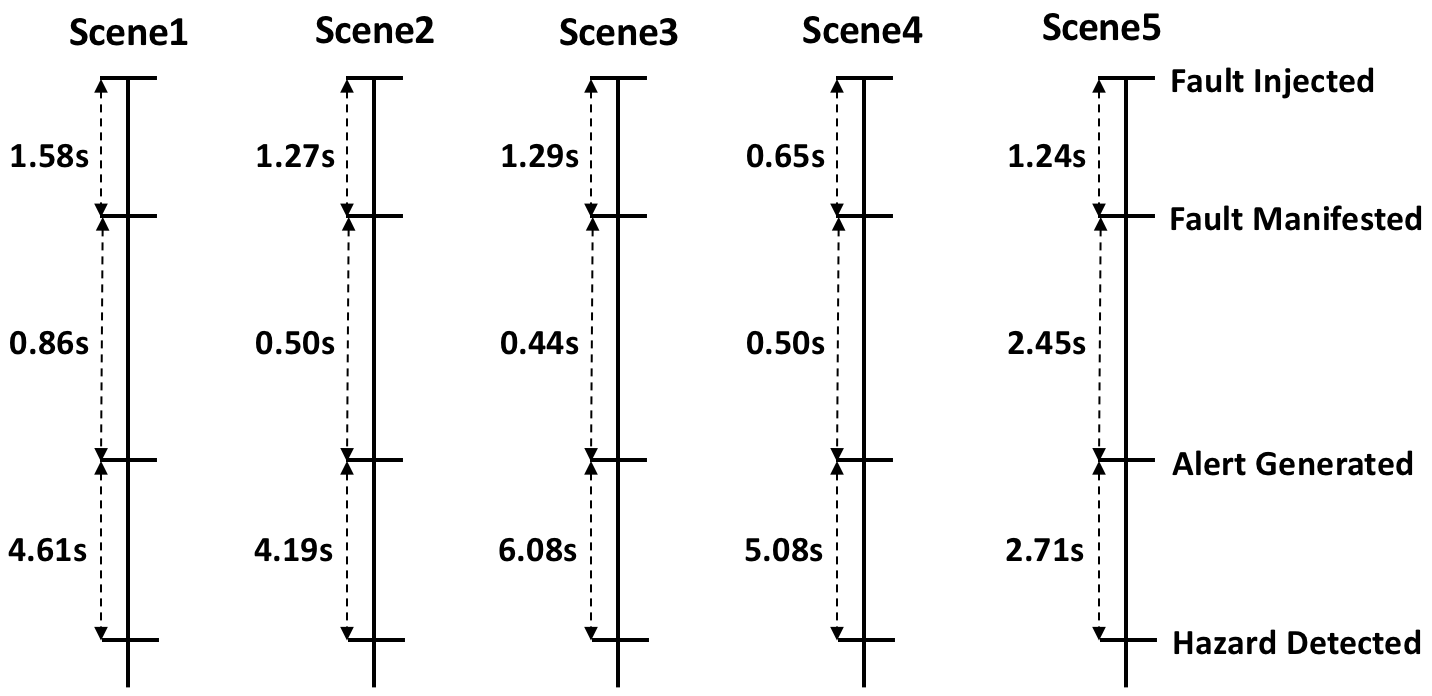}
    \caption{Average fault propagation times for different scenarios.}
    \label{fig:time_graph_guided}
    \vspace{-2em}
\end{figure}

We also analyzed the propagation time of the faults into the system. The manifestation time ($t_m$) of the faults is calculated during post processing of logs, but the hazard ($t_h$) and alert ($t_a$) times are recorded at runtime. All the times are measured with respect to the fault injection time, i.e., after the faults are triggered. The average fault propagation times have been shown in Fig. \ref{fig:time_graph_guided}. %In the cases that the alerts were generated, 
The time between the first alert and the hazard occurrence ($t_r=t_h-t_a$) is calculated as $t_r$, representing the maximum time a human driver will have to take control of the vehicle after seeing an alert. The time a driver actually takes to react to an alert is defined as driver reaction time ($t_R$). A safe AV controller should have a $t_r$ which is greater than the reaction time $t_R$. The reaction time varies from driver to driver. In \cite{Zhang:EvalCAS}, it is stated that average reaction time of the human drivers is around 1.8s if visual warnings are used. Fig. \ref{fig:time_graph_guided} shows that in \textit{openpilot}, average time difference between alerts and hazards varies between 2.71s and 6.08s which is greater than this average reaction time. So, in general human drivers get enough time to react to alerts. But we found that in almost 30.5\% of the hazardous scenarios, no alerts were generated. This means that the human driver needs to be cautious all the time and should not completely rely on the AV to generate alerts in all the scenarios.

\subsubsection{Performance Under Different Faults}
Table \ref{table:haz_coverage} shows the performance of the \textit{openpilot} for different types of faults injected to the inputs of ACC and LKAS systems. As mentioned in section \ref{fault_injection_framework}, four different fault models were considered for the radar module. In total, 714 faults got activated and in almost 28.9\% cases, the faults caused hazards. We found that the AV controller responds better in case of faults injected to the car sensors, as only 21.3\% of cases lead to hazards. 

For the vision module, we injected faults to both input images and output path model. Hazards occurred in cases of images with rain, snow, occlusion, and change of contrast effects. Other effects such as fog, brightness, and blur did not cause any hazards.
%In all these cases, `steer saturated' alert showed up. This alert indicates that the required steer torque is greater than a pre-defined threshold (set inside the openpilot). % for random time experiments, it's not 100% curiously.
Almost all the faults injected to the path model (detected left and right lane positions by the vision module) led to hazardous events (99.8\%). These erroneous values caused the vehicle go out of the lane and in some cases, also collide with the lead vehicle (H1 and H3 together). Faults were also injected to the relative distance values generated from the vision module. But, no hazards happened in these cases because ACC mainly depends on the radar data for perceiving the lead vehicle's status. Injecting faults to both radar and vision modules at the same time caused H1 hazards but that is mainly due to the faults in radar data, indicating that the vision data does not have much effect on ACC outputs.

\begin{table}[!t]
\renewcommand{\arraystretch}{1.3}
\caption{Fault Activation rate and Hazard Coverage for guided vs. random fault injection across different fault types.}
\label{table:haz_coverage}
\centering
\setlength\tabcolsep{5pt}
\begin{tabular}{|c|c|c|c|c|c|}
\hline
\multirow{2}{*}{} & \multirow{2}{*}{} & \multicolumn{2}{|c|}{\tbt{STPA Guided}}  & \multicolumn{2}{|c|}{\tbt{Random}}\\
\cline{3-6}
\makecell{\rotatebox[origin=c]{90}{\makecell{Target\\Module}}} & \makecell{Fault Type} & \makecell{Faults\\Activated\\\# (\%)} & \makecell{Hazard\\Coverage\\\# (\%)} & \makecell{Faults\\Activated\\\# (\%)} & \makecell{Hazard\\Coverage\\\# (\%)} \\
\hline
\multirow{4}{*}{\rotatebox[origin=c]{90}{\tbt{RADAR}}} & \makecell{\tbt{Chaff}\\} & \makecell{432 (90.0)} & \makecell{154 (35.6)} & \makecell{480 (100.0)}& \makecell{260 (54.2)}\\
\cline{2-6}
& \makecell{\tbt{Invisible Vehicle}\\} & \makecell{108 (90.0)} & \makecell{9 (8.3)} & \makecell{120 (100.0)}& \makecell{39 (32.5)}\\
\cline{2-6}
& \makecell{\tbt{Ghost Vehicle}\\} & \makecell{155 (25.0)} & \makecell{34 (21.9)} & \makecell{620 (100.0)}& \makecell{88 (14.2)}\\
\cline{2-6}
& \makecell{\tbt{Radar Jamming}\\} & \makecell{19 (95.0)} & \makecell{9 (47.4)} & \makecell{20 (100.0)}& \makecell{4 (20.0)}\\
\hline
\multirow{2}{*}{\rotatebox[origin=c]{90}{\makecell{Car\\Sensors}}} & \makecell{\tbt{Speed Sensor}\\} & \makecell{138 (57.5)} & \makecell{11 (8.0)} & \makecell{240 (100.0)}& \makecell{15 (6.2)}\\
\cline{2-6}
& \makecell{\tbt{Steer Sensor}\\} & \makecell{190 (95.0)} & \makecell{59 (31.1)} & \makecell{200 (100.0)}& \makecell{77 (38.5)}\\
\hline
\multirow{7}{*}{\rotatebox[origin=c]{90}{\makecell{Vision\\Inputs}}} & \makecell{\tbt{Rain}\\} & \makecell{140 (70.0)} & \makecell{130 (92.9)} & \makecell{200 (100.0)}& \makecell{135 (67.5)}\\
\cline{2-6}
& \makecell{\tbt{Fog}\\} & \makecell{139 (69.5)} & \makecell{0 (0.0)} & \makecell{200 (100.0)}& \makecell{1 (0.5)}\\
\cline{2-6}
& \makecell{\tbt{Snow}\\} & \makecell{140 (70.0)} & \makecell{57 (40.7)} & \makecell{200 (100.0)}& \makecell{56 (28.0)}\\
\cline{2-6}
& \makecell{\tbt{Occlusion}\\} & \makecell{140 (70.0)} & \makecell{72 (51.4)} & \makecell{200 (100.0)}& \makecell{74 (37.0)}\\
\cline{2-6}
& \makecell{\tbt{Contrast}\\} & \makecell{139 (69.5)} & \makecell{78 (56.1)} & \makecell{200 (100.0)}& \makecell{87 (43.5)}\\
\cline{2-6}
& \makecell{\tbt{Brightness}\\} & \makecell{140 (70.0)} & \makecell{0 (0.0)} & \makecell{200 (100.0)}& \makecell{1 (0.5)}\\
\cline{2-6}
& \makecell{\tbt{Blur}\\} & \makecell{126 (70.0)} & \makecell{0 (0.0)}& \makecell{180 (100.0)} & \makecell{1 (0.6)}\\
\hline
\multirow{3}{*}{\rotatebox[origin=c]{90}{\makecell{Vision\\Outputs}}} & \makecell{Camera\\Unavailable\\} & \makecell{18 (90.0)} & \makecell{12 (66.7)} & \makecell{20 (100.0)}& \makecell{6 (30.0)}\\
\cline{2-6}
& \makecell{\tbt{Path Model}\\} & \makecell{540 (90.0)} & \makecell{539 (99.8)} & \makecell{600 (100.0)}& \makecell{470 (78.3)}\\
\cline{2-6}
& \makecell{\tbt{Relative Distance}\\} & \makecell{557 (56.8)} & \makecell{0 (0.0)} & \makecell{980 (100.0)}& \makecell{0 (0.0)}\\
\hline
\rotatebox[origin=c]{90}{\makecell{Radar\\Vision}} & \makecell{\tbt{Relative Distance}\\} & \makecell{557 (56.8)} & \makecell{152 (27.3)} & \makecell{980 (100.0)}& \makecell{320 (32.7)}\\
\hline
%\multicolumn{2}{|c|}{\textbf{Total}} & \makecell{\textbf{3678}\\\textbf{(65.2\%)}} & \textbf{35.78} & \makecell{\textbf{5640}\\\textbf{(100\%)}} & \textbf{28.97} \\
%\hline
\end{tabular}
\vspace{-2em}
\end{table}
\vspace{-0.5em}
\subsection{Observations}
This section summarizes the findings derived from the fault injection experiments in the form of four main observations.

\begin{figure}[!t]
    \centering
    \begin{subfigure}[b]{0.25\textwidth}
        \centering
        \includegraphics[width=\textwidth]{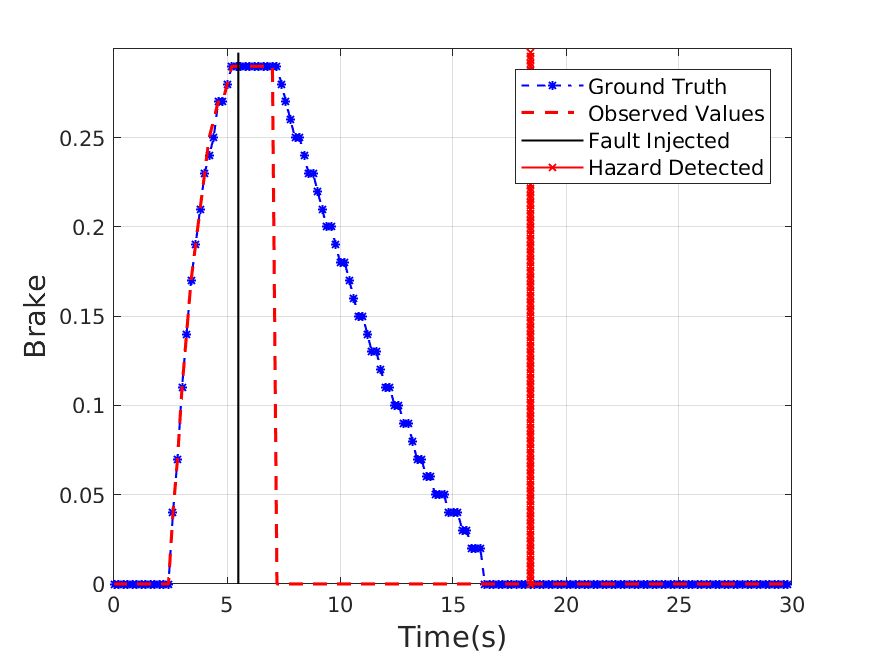}
        \caption{}
        \label{fig:UCAbrake_radj}
        \vspace{-1.5em}
    \end{subfigure}%
    %\hfill%
    \begin{subfigure}[b]{0.25\textwidth}
        \centering
        \includegraphics[width=\textwidth]{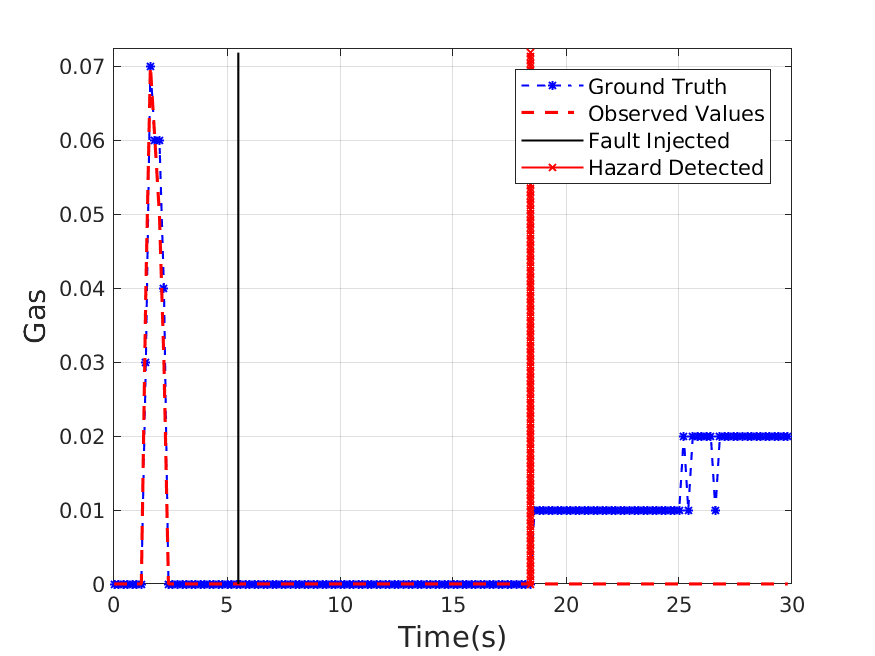}
        \caption{}
        \label{fig:UCAgas_radj}
        \vspace{-1.5em}
    \end{subfigure}
    %\hfill%
    \caption{Both (a) Brake and (b) Gas become unavailable because of a 'radar jamming' fault.}
    \label{fig:UCARadJ}
    \vspace{-2.0em}
\end{figure}

\noindent\fbox{
    \parbox{0.95\linewidth}{
        \textbf{Observation 1:} \textit{Openpilot}, the open source driving agent studied in this work, cannot tolerate safety-critical faults.
    }
}
\begin{itemize}
    \item Table \ref{table:summary} shows that the alert generation rate with respect to the activated faults is 42.7\%. This means that in a large number of cases ($\sim$57.3\%), faults to the inputs degraded the performance of the AV controller, but still went unnoticed.
    \item Although the number of alerts (1570) is larger than the number of hazards (1316), in 10.9\% of cases (with respect to total activated faults) hazards happened but no alerts showed up to warn the driver (Table \ref{table:summary}).
    \item In some cases, both H1 and H3 hazards occurred at the same time (Fig. \ref{fig:haz_dis}). One possible reason was that the vehicle went out of lane because of the faults injected to vision module, then AV controller tried to get it back inside the lane and the vehicle collided with the lead vehicle. Currently, \textit{openpilot} is not equipped with 'lane-change' assist systems, so it is not ready to perfectly handle these kinds of situations.
    \item FCW showed up in the following driving scenarios: 
    \begin{itemize}
        \item Scene2: lead is moving with low speed (25mph).
        \item Scene4: lead slows down first and then  accelerates.
        \item Scene5: lead slows down to a full stop.
    \end{itemize}
    In these cases, average speed of the lead vehicle is low, i.e., relative speed (difference between HV and LV speeds) was high, and, thus, the relative distance was decreasing at a higher rate. So, the FCW was triggered. 
    In the other two scenarios, the average speed of the lead vehicle was higher and no FCW alerts showed up.
\end{itemize}
\noindent\fbox{
    \parbox{0.95\linewidth}{
        \textbf{Observation 2:} The recovery measures in \textit{openpilot} are insufficient for the faults to the input modules.
    }
}
\begin{itemize}
    \item \textit{Openpilot} uses both radar and camera data to calculate relative distance between the lead and host vehicles. But, ACC mainly depends on the radar data, and relative distance coming from vision data only helps to make it more accurate. Table \ref{table:haz_coverage} shows that vision data could not help to avoid collision in case of faulty radar data. When we only injected faults to the relative distance coming from the vision outputs, no hazards happened as the radar data was not perturbed. Injecting faults to both radar and vision data could lead to hazardous scenarios, but they were mainly caused by faulty radar data. So, the technique used by \textit{openpilot} for fusion of the radar and vision data is not sufficient for reducing the risk of collision in case of faulty radar data.
    \item \textit{Openpilot} generates visual and auditory alerts if it detects any problems. Overall only four types of alerts showed up during our experiments, (i) Steer Saturated: if the required steer torque is larger than a threshold, (ii) FCW: if the controller detects that the lead vehicle is getting too close and there is a possibility of collision, (iii) CAN Error: if the RADAR message becomes unavailable, and (iv) Model Error: if the Vision message is unavailable. The warning messages were shown to the driver to take control and avoid unwanted circumstances. Table \ref{table:summary} shows that in 914 out of 1316 cases, hazard occurred after the alerts had been generated. This means that if the driver fails to take control in time, the autonomous controller does not have adequate safety mechanisms to avoid the hazards.
    \item In case of unavailability of data from Radar or Camera, alert messages were displayed to the driver to take control of the vehicle and autonomous control system stops operating (Fig. \ref{fig:UCARadJ}) instead of taking any recovery actions such as slowing down the vehicle. So, the driver must be alert to take control at any time to avoid any hazardous situations.
    \item The lane marker detection algorithm works fine in case of Fog, Brightness, and Blur effects. Very few hazardous circumstances have occurred in these cases according to Table \ref{table:haz_coverage}. But its performance dropped in case of Rain, Snow, Occlusion, and Contrast changes, and thus the system suffered from H3 hazards. This means that the controller did not take sufficient preventive or corrective actions to stop the vehicle from going out of the lane. Although in these cases the \textit{openpilot} generated alerts, those alerts were mainly related to steering saturation.
\end{itemize}

\noindent\fbox{
    \parbox{0.95\linewidth}{
        \textbf{Observation 3:} The time interval between generation of first alert and the occurrence of a hazard is greater than average reaction time of the drivers.
    }
}
\begin{itemize}
    \item Fig. \ref{fig:time_graph_guided} shows that the average time interval between raising an alert and the occurrence of a hazard is always greater than 1.8s (the average reaction time of the drivers). So, in all those cases, the human driver should have enough time to take control of the vehicle.
\end{itemize}

\noindent\fbox{
    \parbox{0.95\linewidth}{
        \textbf{Observation 4:} Context-based fault injection shows relatively better manifestation rate and hazard coverage than the random time based injection.
    }
}
\begin{itemize}
    \item Table \ref{table:summary} shows that in case of STPA guided fault injection, the activation rate is lower than random fault injection. Because in guided injection, the trigger conditions are derived based on the context (Table \ref{tab:context}) which may not always happen depending on the simulated driving scenario. But in random injection, a random time during simulation is chosen as the trigger which will always result in a fault injection (100\% activation rate), i.e., random injection technique practically injects larger number of faults using the same number of experiments. 
    \item Random triggers resulted in a much higher number of injections (5640 vs. 3678) and slightly larger number of hazards (1634 vs. 1316). However, the fault manifestation (99.4\% vs. 93.5\%) and hazard coverage (35.8\% vs. 29.0\%) (with respect to the number of activated faults) were higher in guided fault injection. This observation supports our hypothesis that context based triggers can better simulate the safety-critical scenarios by activating faults only at critical times during simulation. On the other hand, a randomly selected injection time may not always give the activated faults enough chance to further propagate and cause safety-critical impacts.
\end{itemize}

From the above discussion, we can conclude that the \textit{openpilot} shows low resilience to the changes in environment and faults affecting sensor data. The driver must always be alert to take control and is still responsible for most of the safety-critical functions of the vehicle and monitoring of the environment. This is consistent with the conclusions made by \cite{subho2018hands} that AVs need better situational awareness to be able to preemptively avoid accidents in a timely fashion.

\vspace{-0.5em}
\section{Related Work}
The related work on resilience assessment of AVs can be classified into three categories: (i) Hazard and requirements analysis, (ii) Fault injection testing, and (iii) Adversarial ML.

\textbf{Hazard and requirements analysis:}
%A powerful hazard analysis technique is required for a complex cyber-physical system like AV to identify the potential hazards and suggest safety requirements at the very beginning of the development process. STPA has been proven to be a powerful and dependable hazard analysis technique for different safety critical systems. But, very few efforts have been made so far to utilize STPA in AV systems. 
The international safety standard for automobiles (ISO 26262) recommends identifying potential safety hazards and defining safety requirements to implement mechanisms that can detect and mitigate hazards. Commonly used hazard analysis techniques such as Failure Mode and Effects Analysis (FMEA) \cite{segismundo2008failure} \cite{mader2011computer} and Fault Tree Analysis (FTA) \cite{lambert2004use} primarily focus on probabilistic analysis of individual component failures in the system. Other causal factors, e.g., complex software errors and unsafe interactions among components given underlying system context, are often not thoroughly studied in these techniques. To overcome these limitations, STPA \cite{leveson2011engineering}, a technique based on STAMP (Systems-Theoretic Accident Model and Processes), has been proposed. STPA models the accidents as complex dynamic processes resulting from inadequate control mechanisms that violate safety constraints. STPA has been previously applied to hazard analysis in AVs \cite{Haneet:LKAS} \cite{Abdulkhaleq:STPAav}. \cite{STPA2SISystems:Abdulkhaleq} and \cite{STPA2CAS:Sulaman} used STPA for hazard analysis of collision avoidance systems. Asim et al. \cite{abdulkhaleq2016systematic} proposed a semi-autonomous method for test case generation from  manually generated STPA tables. In this work, we use STPA to analyze the potential unsafe control actions in LKAS and ACC modules of an AV, and use the system context for unsafe control actions to identify the critical triggers for injecting faults into the control software.
%\todo[inline]{cite Asim's work which is cited in the proposal--cited} \todo[inline]{cite SafeComp where we used STPA and fault injection for surgical robots--already cited in the first line of this paragraph}
% Abdulkhaleq et al. \cite{STPA2DepArch:Abdulkhaleq} proposed an STPA based systematic approach to evaluate and develop a dependable architecture for fully AV system. They mainly focused on the application of STPA in architecture level by dividing the complete functional design in three abstraction layers: automated driving level, system level, and component level. Difficulty in linking different control modules at multiple levels may limit the completeness of the hazards.

\textbf{Safety validation by software fault injection:} The ISO 26262 standard also emphasizes the importance of fault-injection testing \cite{Cotroneo:finject} to evaluate the performance of safety mechanisms in automotive systems. Software fault injection and fuzz testing have been previously used for the resilience and robustness testing of automotive software and AVs \cite{jha2018avfi}\cite{koopman2017autonomous}\cite{hutchison2018robustness}.
\cite{jha2018avfi} presented preliminary results on a fault injector for evaluating the end-to-end resilience of AVs to faults in the sensor inputs, neural networks, and hardware software components of an AV simulator. In \cite{hutchison2018robustness}, fuzz testing was used to find bugs in 17 different autonomous systems by injecting exceptional values from a fault dictionary into the interface messages at random periods during system run-time. In our approach the locations and trigger conditions for injections are driven by the STPA hazard analysis and the overall resilience of system is assessed by measuring the manifestation of unsafe scenarios, hazard coverage, and fault propagation times.

%of dependability of different systems, such as smart power grid \cite{Faza:powergrid} \cite{Tseng:powergrid}, operating systems \cite{Kropp:posixOS} \cite{Chen:OS}, cloud platforms \cite{Martino:Saas} \cite{Pham:cloudVal}, and networks \cite{Stott:network}. In this work, a compile-time fault injection strategy similar to \cite{Alemzadeh:Safecomp} has been followed. 

\textbf{Adversarial Machine Learning:} Another active area of research focuses on testing ML algorithms used for computer vision, perception, and planning in AVs \cite{evtimov2017robust},\cite{Pei:Deepxplore}\cite{DeepTest:Tian}. The closest work in this category to us is DeepTest \cite{DeepTest:Tian} that automatically tests the DNN-driven steering angle prediction algorithms (LKAS feature) in AVs by injecting single (e.g. contrast, brightness, blur etc.) or composite (e.g. rain and fog) faults into the input images. %They mainly focused on testing the lateral control  of the vehicle using perturbed images, and observing the steering angle derived by the DNN algorithm. 
The end-to-end resilience assessment of AV, the impact of faults on the performance of ACC system, and their propagation into causing hazards were not studied by DeepTest. In this paper, we simulated the real-world effects of environmental conditions (e.g. rain, snow, fog) as well as sensor faults on the performance of both ACC and LKAS systems. We not only tested the impact of faulty images on the performance of vision algorithms, but also characterized the propagation of the faults in the control software of the driving agent and the possibility of causing hazards.
\vspace{-0.5em}
\section{Conclusion}
We presented a fault injection framework for evaluating the resilience of an open-source AV software (\textit{openpilot}) in the face of environmental conditions and faults affecting car sensors. We presented a strategic fault injection approach where the targets and triggers for injecting the faults are derived from the unsafe control actions and critical system context identified during the hazard analysis of the system. %STPA hazard analysis technique is used to identify the safety hazard scenarios and their potential causal factors. A software fault injection engine is developed to inject faults to the Vision and RADAR modules and car sensors. %To mimic the impact of real-world environmental conditions such as rain, fog, and occlusion, synthesized images have been fed to a lane marker detection algorithm used as the vision module of the PC simulator of \textit{openpilot}. 
The experimental results indicate that: i) our strategic fault injection approach provides a better coverage of hazard scenarios compared to random fault injection, and ii) although \textit{openpilot} is equipped with safety mechanisms to alert the driver, faulty sensor data might still cause it to issue unsafe control actions, leading to safety hazards. Also, a non-negligible percentage of those hazards might go undetected by the controller. These results indicate the need for developing more effective situational awareness and safety mechanisms in \textit{openpilot} controller. Further evaluation of the proposed fault injection approach with a wider set of scenarios and using other AV controllers and simulators is the subject of future work. 
\vspace{-0.5em}
%\subsection*{Future Work}
\begin{comment}In this work, image processing and shallow neural network based vision algorithm have been used. There are several deep learning based techniques for lane detection and predicting steer angle. We plan to use a similar deep neural network based technique in the vision module. We are also working on a safety monitoring system that can detect the erroneous behavior and take recovery measures before the hazard happens.
\end{comment}
%\appendices
%\blindtext

%\section*{Acknowledgment}

\ifCLASSOPTIONcaptionsoff
  \newpage
\fi
\bibliographystyle{IEEEtran}
\bibliography{IEEEabrv,OpenPilot}

\begin{thebibliography}{10}
\providecommand{\url}[1]{#1}
\csname url@samestyle\endcsname
\providecommand{\newblock}{\relax}
\providecommand{\bibinfo}[2]{#2}
\providecommand{\BIBentrySTDinterwordspacing}{\spaceskip=0pt\relax}
\providecommand{\BIBentryALTinterwordstretchfactor}{4}
\providecommand{\BIBentryALTinterwordspacing}{\spaceskip=\fontdimen2\font plus
\BIBentryALTinterwordstretchfactor\fontdimen3\font minus
  \fontdimen4\font\relax}
\providecommand{\BIBforeignlanguage}[2]{{%
\expandafter\ifx\csname l@#1\endcsname\relax
\typeout{** WARNING: IEEEtran.bst: No hyphenation pattern has been}%
\typeout{** loaded for the language `#1'. Using the pattern for}%
\typeout{** the default language instead.}%
\else
\language=\csname l@#1\endcsname
\fi
#2}}
\providecommand{\BIBdecl}{\relax}
\BIBdecl

\bibitem{TeslaIncident2016}
A.~Singhvi and K.~Russell, ``Inside the self-driving tesla fatal accident,''
  https://www.nytimes.com/interactive/2016/07/01/business/inside-tesla-accident.html,
  accessed: 2018-05-11.

\bibitem{TeslaIncident2018}
N.~Boudette, ``Fatal tesla crash raises new questions about autopilot system,''
  https://www.nytimes.com/2018/03/31/business/tesla-crash-autopilot-musk.html,
  accessed: 2018-05-11.

\bibitem{UberIncident}
A.~Efrati, ``Uber finds deadly accident likely caused by software set to ignore
  objects on road,''
  https://www.theinformation.com/articles/uber-finds-deadly-accident-likely-caused-by-software-set-to-ignore-objects-on-road,
  accessed: 2018-05-08.

\bibitem{evtimov2017robust}
I.~Evtimov \emph{et~al.}, ``Robust physical-world attacks on deep learning
  models,'' \emph{arXiv preprint arXiv:1707.08945}, vol.~1, 2017.

\bibitem{Pei:Deepxplore}
K.~Pei \emph{et~al.}, ``Deepxplore: Automated whitebox testing of deep learning
  systems,'' in \emph{Proceedings of the 26th Symposium on Operating Systems
  Principles}.\hskip 1em plus 0.5em minus 0.4em\relax ACM, 2017, pp. 1--18.

\bibitem{DeepTest:Tian}
Y.~Tian \emph{et~al.}, ``Deeptest: Automated testing of
  deep-neural-network-driven autonomous cars,'' pp. 303--314, 2018.

\bibitem{RadarFault:Petit}
J.~Petit and S.~E. Shladover, ``Potential cyberattacks on automated vehicles,''
  \emph{IEEE Transactions on Intelligent Transportation Systems}, vol.~16,
  no.~2, pp. 546--556, 2015.

\bibitem{openpilot:Git}
``Openpilot git repo,'' https://github.com/commaai/openpilot.

\bibitem{subho2018hands}
S.~S. Banerjee \emph{et~al.}, ``Hands off the wheel in autonomous vehicles? a
  systems perspective on over a million miles of field data,'' in \emph{2018
  48th Annual IEEE/IFIP International Conference on Dependable Systems and
  Networks (DSN)}.\hskip 1em plus 0.5em minus 0.4em\relax IEEE, 2018, pp.
  586--597.

\bibitem{leveson2011engineering}
N.~Leveson, \emph{Engineering a safer world: Systems thinking applied to
  safety}.\hskip 1em plus 0.5em minus 0.4em\relax MIT press, 2011.

\bibitem{comma.ai}
``Comma.ai,'' https://comma.ai/, accessed: 2018-05-12.

\bibitem{openpilot:HW}
``Eon dashcam devkit,'' https://shop.comma.ai/products/eon-dashcam-devkit,
  accessed: 2018-05-12.

\bibitem{DBLP:journals/corr/SantanaH16}
\BIBentryALTinterwordspacing
E.~Santana and G.~Hotz, ``Learning a driving simulator,'' \emph{CoRR}, vol.
  abs/1608.01230, 2016. [Online]. Available:
  \url{http://arxiv.org/abs/1608.01230}
\BIBentrySTDinterwordspacing

\bibitem{Vision:HOG}
N.~Dalal and B.~Triggs, ``Histograms of oriented gradients for human
  detection,'' in \emph{Computer Vision and Pattern Recognition, 2005. CVPR
  2005. IEEE Computer Society Conference on}, vol.~1.\hskip 1em plus 0.5em
  minus 0.4em\relax IEEE, 2005, pp. 886--893.

\bibitem{bishop1995neural}
C.~Bishop \emph{et~al.}, \emph{Neural networks for pattern recognition}.\hskip
  1em plus 0.5em minus 0.4em\relax Oxford university press, 1995.

\bibitem{lane:dataset}
M.~Aly, ``Real time detection of lane markers in urban streets,'' in
  \emph{Intelligent Vehicles Symposium, 2008 IEEE}.\hskip 1em plus 0.5em minus
  0.4em\relax IEEE, 2008, pp. 7--12.

\bibitem{Sobel:HGrad}
I.~Sobel, ``An isotropic 3$\times$ 3 image gradient operator,'' \emph{Machine
  vision for three-dimensional scenes}, pp. 376--379, 1990.

\bibitem{Ester:DBSCAN}
M.~Ester \emph{et~al.}, ``A density-based algorithm for discovering clusters in
  large spatial databases with noise.'' in \emph{Kdd}, vol.~96, no.~34, 1996,
  pp. 226--231.

\bibitem{openpilotVis:Git}
https://github.com/UVA-DSA/OpenpilotWithVision.

\bibitem{Alemzadeh:Safecomp}
H.~Alemzadeh \emph{et~al.}, ``Systems-theoretic safety assessment of robotic
  telesurgical systems,'' in \emph{Proceedings of the 34th International
  Conference on Computer Safety, Reliability, and Security - Volume 9337}, ser.
  SAFECOMP 2015, 2015, pp. 213--227.

\bibitem{RainStreak:Garg}
K.~Garg and S.~K. Nayar, ``Photorealistic rendering of rain streaks,'' in
  \emph{ACM Transactions on Graphics (TOG)}, vol.~25, no.~3.\hskip 1em plus
  0.5em minus 0.4em\relax ACM, 2006, pp. 996--1002.

\bibitem{RadarFault:Rasshofer}
R.~Rasshofer \emph{et~al.}, ``Influences of weather phenomena on automotive
  laser radar systems,'' \emph{Advances in Radio Science: ARS}, vol.~9, p.~49,
  2011.

\bibitem{Zhang:EvalCAS}
Y.~Zhang \emph{et~al.}, ``A new threat assessment measure for collision
  avoidance systems,'' in \emph{Intelligent Transportation Systems Conference,
  2006. ITSC'06. IEEE}.\hskip 1em plus 0.5em minus 0.4em\relax IEEE, 2006, pp.
  968--975.

\bibitem{segismundo2008failure}
A.~Segismundo and P.~Augusto Cauchick~Miguel, ``Failure mode and effects
  analysis (fmea) in the context of risk management in new product development:
  A case study in an automotive company,'' \emph{International Journal of
  Quality \& Reliability Management}, vol.~25, no.~9, pp. 899--912, 2008.

\bibitem{mader2011computer}
R.~Mader \emph{et~al.}, ``Computer-aided pha, fta and fmea for automotive
  embedded systems,'' in \emph{International Conference on Computer Safety,
  Reliability, and Security}.\hskip 1em plus 0.5em minus 0.4em\relax Springer,
  2011, pp. 113--127.

\bibitem{lambert2004use}
H.~E. Lambert, ``Use of fault tree analysis for automotive reliability and
  safety analysis,'' SAE Technical Paper, Tech. Rep., 2004.

\bibitem{Haneet:LKAS}
H.~S. Mahajan \emph{et~al.}, ``Application of systems theoretic process
  analysis to a lane keeping assist system,'' \emph{Reliability Engineering \&
  System Safety}, vol. 167, pp. 177--183, 2017.

\bibitem{Abdulkhaleq:STPAav}
A.~Abdulkhaleq \emph{et~al.}, ``A systematic approach based on stpa for
  developing a dependable architecture for fully automated driving vehicles,''
  \emph{Procedia Engineering}, vol. 179, pp. 41--51, 2017.

\bibitem{STPA2SISystems:Abdulkhaleq}
A.~Abdulkhaleq and W.~Stefan, ``Experiences with applying stpa to
  software-intensive systems in the automotive domain,'' 01 2013.

\bibitem{STPA2CAS:Sulaman}
S.~Sulaman \emph{et~al.}, ``Hazard analysis of collision avoidance system using
  stpa,'' 05 2014.

\bibitem{abdulkhaleq2016systematic}
A.~Abdulkhaleq and S.~Wagner, ``A systematic and semi-automatic safety-based
  test case generation approach based on systems-theoretic process analysis,''
  \emph{arXiv preprint arXiv:1612.03103}, 2016.

\bibitem{Cotroneo:finject}
D.~Cotroneo and R.~Natella, ``Fault injection for software certification,''
  \emph{IEEE Security Privacy}, vol.~11, no.~4, pp. 38--45, July 2013.

\bibitem{jha2018avfi}
``Avfi: Fault injection for autonomous vehicles,'' in \emph{2018 48th Annual
  IEEE/IFIP International Conference on Dependable Systems and Networks
  Workshops (DSN-W)}.\hskip 1em plus 0.5em minus 0.4em\relax IEEE, 2018, pp.
  55--56.

\bibitem{koopman2017autonomous}
P.~Koopman and M.~Wagner, ``Autonomous vehicle safety: An interdisciplinary
  challenge,'' \emph{IEEE Intelligent Transportation Systems Magazine}, vol.~9,
  no.~1, pp. 90--96, 2017.

\bibitem{hutchison2018robustness}
C.~Hutchison \emph{et~al.}, ``Robustness testing of autonomy software,'' in
  \emph{Proceedings of the 40th International Conference on Software
  Engineering: Software Engineering in Practice}.\hskip 1em plus 0.5em minus
  0.4em\relax ACM, 2018, pp. 276--285.

\end{thebibliography}
%\begin{IEEEbiography}[{\includegraphics[width=1in,height=1.25in,clip,keepaspectratio]{picture}}]{John Doe}
%\blindtext
%\end{IEEEbiography}
\end{document}